\documentclass{article}
\usepackage[utf8]{inputenc}
\usepackage{amsmath, amssymb, graphicx}
\usepackage[margin=1in]{geometry}
\usepackage{scrextend}
\usepackage[usenames, dvipsnames]{color}
\usepackage{bbm,bm}
\usepackage{mathtools}
\usepackage{comment}
\usepackage[table,xcdraw]{xcolor}
\usepackage[semicolon, authoryear]{natbib}
\usepackage[hyperfootnotes = true,colorlinks,citecolor=blue,urlcolor=blue]{hyperref}
\usepackage{caption}
\usepackage{subcaption}
\usepackage{amsthm}

\usepackage{fancyhdr}

\usepackage{abstract}

\usepackage{setspace}
\newcommand\pk{p_{\text{keep}}}
\newcommand\pn{p_{\text{new}}}
\newcommand\ak{a_{\text{keep}}}
\newcommand\an{a_{\text{new}}}

\newcommand{\footremember}[2]{%
    \footnote{#2}
    \newcounter{#1}
    \setcounter{#1}{\value{footnote}}%
}
\newcommand{\footrecall}[1]{%
    \footnotemark[\value{#1}]%
}

\title{Clustering Future Scenarios Based on Predicted Range Maps}
\author{Matthew Davidow\footremember{x}{Corresponding author: Matthew Davidow, Email address: \url{mbd83@cornell.edu}}\footremember{b}{Department of Statistics and Data Science, Cornell University}
\and Toryn L. J. Schafer\footrecall{b} 
\and Cory Merow\footremember{a}{Eversource Energy Center and Department of Ecology and Evolutionary Biology, University of Connecticut}
\and Judy Che-Castaldo\footremember{c}{Department of Conservation and Science, Lincoln Park Zoo}
\and Marie-Christine D\"{u}ker\footrecall{b} 
\and Emily Feng\footrecall{c}
\and David S. Matteson\footrecall{b}}
\date{\today}

\begin{document}

\maketitle

\begin{abstract}
% abstract points supposed to be 
% Point 1: set the context for and purpose of the work;
% Point 2: indicate the approach and methods;
% Point 3: outline the main results;
% Point 4: identify the conclusions and the wider implications.
% see Journal guidelines, included in the JournalGuidelines folder in this overleaf project,
% which is also found at https://besjournals.onlinelibrary.wiley.com/hub/journal/2041210X/author-guidelines
\noindent
\vspace{-1cm}

\begin{enumerate}
\item Predictions of biodiversity trajectories under climate change are crucial in order to act effectively in maintaining the diversity of species. In many ecological applications, future predictions are made under various global warming scenarios as described by a range of different climate models. We propose a clustering methodology to synthesize and interpret the outputs of these various predictions.
\item We propose an interpretable and flexible two step methodology to measure the similarity between predicted species range maps and to cluster the future scenario predictions utilizing a spectral clustering technique.
\item  We find that clustering based on predicted species range maps is mainly driven by the amount of warming rather than climate model or future scenario. We contrast this with clustering based only on predicted climate variables, which is driven primarily  by climate models, i.e., scenarios of the same climate model are clustered together, even when the amount of warming input to the models is varied.
\item The differences between species-based and climate-based clusterings illustrate that it is crucial to incorporate ecological information to understand the relevant differences between climate models. Our findings can be used to better synthesize forecasts of biodiversity change under the wide spectrum of results that emerge when considering potential future scenarios.
%  Not all areas of the globe and aspects of climate are equally important in determining ecological outcomes.
\end{enumerate}
\textbf{Key-words:} biodiversity; clustering; similarity measures; future scenarios; climate change; GCM.
\end{abstract}
% Cory suggestion for data availability:
% github repo public on acceptance of the paper

\section{Introduction}
Natural and human-induced global climate change undeniably threatens the biodiversity of plants and animals and is expected to lead to large-scale range loss \citep{lovejoy2006climate,mackey2008climate,dantas2015climate,radchuk2019adaptive}. A common response of species to global climate change is dispersal or range map shifting \citep{pecl2017biodiveresity} or range map contraction. Understanding the extent of potential ecological responses to a rapidly changing climate is essential to enact effective conservation policies \citep{parry2007climate,hannah2013global,hannah202030}.

Forecasting species distributions or range maps under various scenarios of future climate provides a suite of potential outcomes \citep{burrows2014geographical,jones2015multi,molinos2016climate}. Given a model for species range maps based on current climate, a suite of forecasts can be made using future climate projections such as those from the Coupled Model Intercomparison Project 6 (CMIP6). A comprehensive look at the forecasts under various scenarios for a large number of species can quickly become burdensome to summarize both within and among species. Clustering methods provide a mechanism for extracting lower dimensional information; clustering species range maps across scenarios elicits patterns of responses to climate change within a species and across species.  

Principal component analysis (PCA) is a widely used tool in ecological applications to interpret high-dimensional data \citep{pearson1901liii,jolliffe1986principal,huettmann2001using,janvzekovivc2012pca,Wu}. However, PCA has significant drawbacks in the nonlinear, multivariate setting of the current application. PCA has a tight relationship with correlation which classically fails to capture nonlinear relationships. Secondly, it is not clear how to incorporate information from multiple species into a PCA-based approach, since different species have different sized range maps, and thus different species will naively vary in the principal component contributions due to size alone, which may not be desired. In contrast, we present a flexible alternative, spectral clustering, allowing for any similarity or distance between range maps. Spectral clustering is ideal for our dataset because of the high-dimensional range maps and the desire to amalgamate information from across species. Spectral clustering is a technique to cluster observations given only pairwise similarity or distance measures between the observations. It is well suited for this task because we can define a function that captures our desired notion of pairwise similarity measure between potential range maps, which will be discussed in Section \ref{se:Weights}.

A popular alternative summarisation to clustering is a metric of change in species richness due to climate change known as climate velocity \citep{loarie2009velocity} which can be used as a proxy for net effects on biodiversity change, particularly when detailed biodiversity information is not available \citep{burrows2014geographical,jones2015multi,molinos2016climate}. However, these climate velocities rely entirely on climate information while ignoring ecological data. Such ecological information, such as species' exposure to climate conditions not found in the species' current niches that are described by species range maps models, are important factors to predict the species potential future range maps \citep{trisos2020projected}. We propose to cluster the forecasted species range maps to interpret and investigate the similarities and differences among predictions across different warming scenarios, global climate models (GCMs), and species. We analyze a dataset of $1101$ animals over $34$ different future scenarios, for a total of over $37,000$ range maps. Ecological data informs the predicted range maps, and thus the spectral clustering results will make full use of the available ecological information. In order to assess the amount of information gained from the species range map model, we additionally apply spectral clustering to the scenarios based only on the predicted values of the climate variables. We hypothesize that clustering the scenarios based on these predicted range maps will produce significantly different results than clustering based only on the climatic variables of the scenarios. The differences will emphasize the value of monitoring species specific changes.

Our methodology can be used to cluster based on individual species, on all species together, or on subsets of species such as those species most at risk. Clustering the predicted range maps for a single species across the suite of climate predictions is important to understand the climatic response for the single species, whereas clustering based on all species has the benefit of obtaining a broad understanding of the overarching patterns. Furthermore, clustering can be based only on a subset of the species such as those considered most at risk; summarizing predictions for species most at risk highlights climatic factors that drive high loss of biodiversity.

We first describe prior modelling work to predict the future potential range maps given climate information (Sections \ref{se:2.1} and \ref{se:clim}). Then we discuss our methodology for clustering ecological and climate scenarios (Section \ref{se:Methodology}); first based on predicted range maps, and secondly based on predicted climate variables. Additionally, we provide an illustrating example. In Section \ref{se:resulstanddis}, we discuss the scenario clusterings, with emphasis on the difference between the climate-based and ecological based clustering. We then demonstrate that we have detected meaningful clusters, with range maps within clusters similar to each other, but different than range maps in other clusters. Finally, we make some concluding remarks in Section \ref{se:conclusion}.

% I do not know where this fits: 
%. The future climate scenarios differ in their underlying climate variables.
%Predicted climate scenarios have been clustered with respect to those climate variables by averaging the variables over global regions \citep{giorgi2000uncertainties,cannon2015selecting}. However, spatially averaging the climate variables over global regions loses significant information about the spatial variability of the variables.

\section{Materials and Methods} \label{se:Methods} % 

\subsection{Data: Range Map Predictions} \label{se:2.1}

We used inhomogeneous Poisson point process models (PPPM) \citep{merow2017integrating,renner2015point,warton2010poisson} to predict the future range maps of 1101 terrestrial mammal species in 2070 based on current day presence data and climatic covariates. For current day climatic conditions, five climate variables with minimum pairwise correlation were chosen from a set of 19 commonly available Bioclim variables in the world climate database, WorldCLIM2 \citep{fick2017worldclim}. 

The mean values from 1970-2000 of the five chosen climate variables were used to fit a PPPM for the species occurrence. Occurrence data for each species were obtained from \cite{GBIF:2020}, and PPPMs were fit for species with at least 10 unique presence cells on a global 10km grid. The fitted model was used to predict future potential distributions based on the CMIP6 predictions of the five climate variables.  Predicted relative occurrence rate was converted to binary range maps of presence/absence with a chosen threshold based on the 5\% quantile of predicted relative occurrence rate values of training presences. This approach was used to make predictions for 1101 mammals with sufficient data using the 34 different sets of predicted climate variables. An example of predicted range maps for the Australian sandy inland mouse (\textit{Pseudomys hermannsburgensis}) is shown in Figure \ref{fig:weighting:scenario}.

\subsection{Data: Climate Models}\label{se:clim}

 Climate variables used for prediction are collected in the Coupled Model Intercomparison Project 6 (CMIP6). CMIP6 provides multiple climate predictions which vary by the underlying global climate model (GCM) and the representative concentration pathway (RCP) used. RCPs describe a predicted timeline of greenhouse gas concentrations \citep{eyring2016overview}. Four RCPs are included in CMIP6, which vary in the quantity of greenhouse gas emissions to capture the uncertainty of future emissions. We refer to the RCP trajectory of least emission as the ``optimistic" trajectory,  and refer to the most pessimistic trajectory as the ``extreme'' trajectory. See also Table \ref{tab:ScenTable} for more information about the different climate models. In this work we refer to a \emph{scenario} as a (GCM, RCP) pair; a scenario represents uncertainty both in the evolution of climate, and in future greenhouse gas emissions. There is a variety of publications which studied the climate predictions considered in this paper; see 
\cite{gmd-9-1937-2016,gmd-11-1033-2018,gmd-13-2149-2020,gmd-13-3203-2020}.

% Methodology is not a full section
\subsection{Clustering Methodology}\label{se:Methodology}
First, we introduce some notation regarding the range maps (Section \ref{subsubse1:Methodology}) followed by a way to quantify similarity between range maps (Section \ref{se:Weights}). Furthermore, we utilize spectral clustering to cluster scenarios based on the proposed similarity measure (Section \ref{subsubse3:Methodology}). We conclude this section with a procedure to cluster scenarios based on different climate variables (Section \ref{se:climatebased}), which can be compared to clustering based on the ecologically informed range maps which highlights the importance of utilizing ecological information.

\subsubsection{Range Map Notation} \label{subsubse1:Methodology}
For notational simplicity we focus the presentation of the methodology for a single species. Furthermore, we suppose our range maps are on a $\bm{r}$ by $\bm{c}$ regular grid represented as $(r,c): r=1,\ldots,\bm{r};\ c= 1,\ldots,\bm{c} $. We denote $B_s$ as the binary (presence/absence) predicted range map according to scenario $s$, to be more precise $B_s(r,c) = 1$ if the cell at $(r,c)$ represents a presence, and $B_s(r,c) = 0$ otherwise. It is important to consider how these range maps differ from the present day;  for this reason, we denote by $P$ the present day map. Similarly as for $B_s(r,c)$, we define $P(r,c) = 1$ if the cell at $(r,c)$ is presently occupied, and $0$ otherwise. We define a candidate set $A$ where presences and valid absences may occur, so that non-meaningful cells such as those in the ocean have no effect on quantifying similarity between range maps described in the next section. This binary candidate map $A$ has size $\bm{r}$ by $\bm{c}$, and could for example take the value 1 only when there is land, or alternatively the geographic area reachable from the species considering physical constraints on dispersal such as mountain ranges. We have chosen to take $A$ as the union of the original range map and all scenarios, thus $A(r,c) = 1$ if $P(r,c) = 1$, or there is at least one scenario $s$ such that $B_s(r,c) = 1$. More generally, $A$ is a mask for the $\bm{r}$ by $\bm{c}$ rectangular grid that is either occupied in the present or occupied in at least one future prediction. 

\subsubsection{Quantifying Similarity}\label{se:Weights}

We choose to quantify the similarity (or differences) among the climate scenarios, $s$, with respect to changes in presence/absence from the present day, $P$. For each scenario $s$, we construct a corresponding weighted map, $W_s$, that indicates the nature of the changes. The value of $W_s(r,c)$ is 0 if $A(r,c) = 0$, but when $A(r,c) = 1$, the value of $W_s(r,c)$ depends on the intersection of the value for the present day map and the value under scenario $s$ at location $(r,c)$. The four possible values of $W_s(r,c)$ are shown in Table \ref{tab:WTable}. 

\begin{table}[hbt]
    \centering
    \begin{tabular}{c|c c}
            %predicted $B(r,c)$ & current $P(r,c)$ & \\ 
            & $P(r,c) = 1$ & $P(r,c) = 0$  \\
            \hline
            $B(r,c) = 1$ &  $\pk{}$ & $\pn{}$ \\
            $B(r,c) = 0$ & $\an{}$ & $\ak{}$
    \end{tabular}
    \caption{\label{tab:WTable} Cell weighting values for $W(r,c)$ given $A(r,c) = 1$.}
\end{table}
The four possible cases represent unchanged (kept) presences, $\pk{}$, new presences, $\pn{}$, new absences, $\an{}$, and unchanged absences, $\ak{}$, respectively, where ``keep" and ``new" are with respect to the present day range map, $P$. For example a grid cell $(r,c)$ corresponding to $\pk{}$ is a grid cell that is a presence in the current day range map, $P(r,c) = 1$, and remains a presence according to scenario $s$, $B_s(r,c) = 1$.

% Alternative weighting considered:
The best choice of weightings, ($\pk{}, \pn{}, \an{}, \ak{} $), is dependent on the pairwise similarity metric used for clustering. We choose to compute the pairwise similarity between scenario range maps as the cosine similarity of their weighted range maps, $\text{CosineSimilarity}(W_s,W_{s'})$ or $\text{CS}(W_s,W_{s'})$ for short. We define the cosine similarity between two matrices $W_s,W_{s'}$ of dimension $\bm{r} \times \bm{c}$ as 
\begin{equation}\label{eq:cosSim} 
    \text{CS}(W_s,W_{s'}) = \frac{W_s\cdot W_{s'}}{\| W_s\|_2 \| W_{s'}\|_2 } = \frac{\sum\limits_{r=1}^{\bm{r}}\sum\limits_{c=1}^{\bm{c}} W_s(r,c)W_{s'}(r,c)}{\| W_s\|_2 \| W_{s'}\|_2 }
    \hspace{0.2cm} \text{ with } \hspace{0.2cm}
    \| W_s\|_2  = \Big(\sum\limits_{r=1}^{\bm{r}}\sum\limits_{c=1}^{\bm{c}} W_s(r,c)^2\Big)^{\frac{1}{2}}.
\end{equation}
The product in the numerator in \eqref{eq:cosSim} suggests that when $W_s(r,c)$ and $W_{s'}(r,c)$ are zero there is no contribution to the measure of similarity suggesting that the weighting scheme, ($\pk{}, \pn{}, \an{}, \ak{} $),  should be constructed such that meaningful contributions to similarity have non-zero values.

\begin{figure}
    \begin{subfigure}{.31\textwidth}
        \centering
        \includegraphics[width=\linewidth]{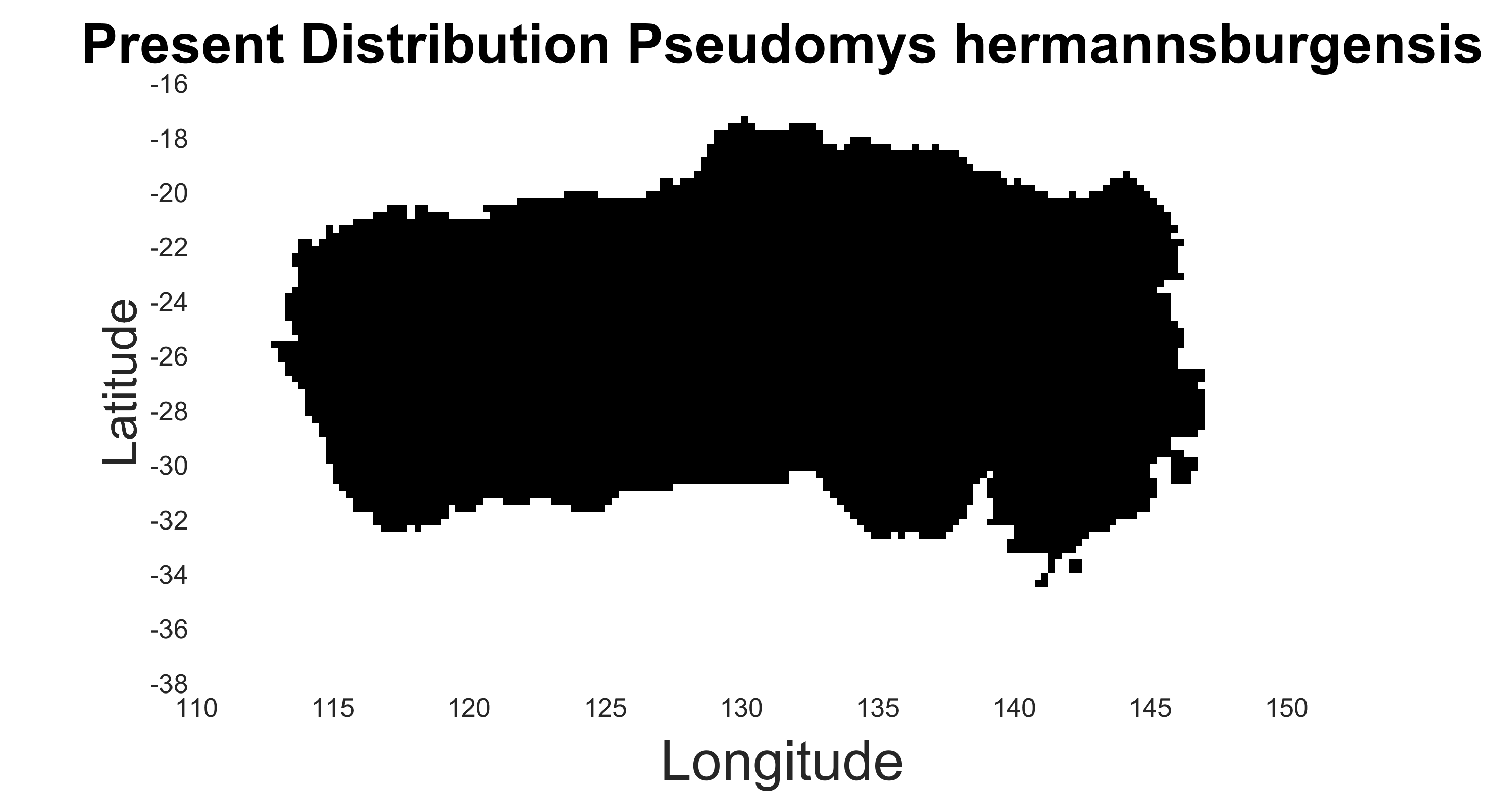}
        \caption{Present Day Range Map}
        \label{fig:weighting:present}
    \end{subfigure}
    \hfill
    \begin{subfigure}{.31\textwidth}
        \centering
        \includegraphics[width=\linewidth]{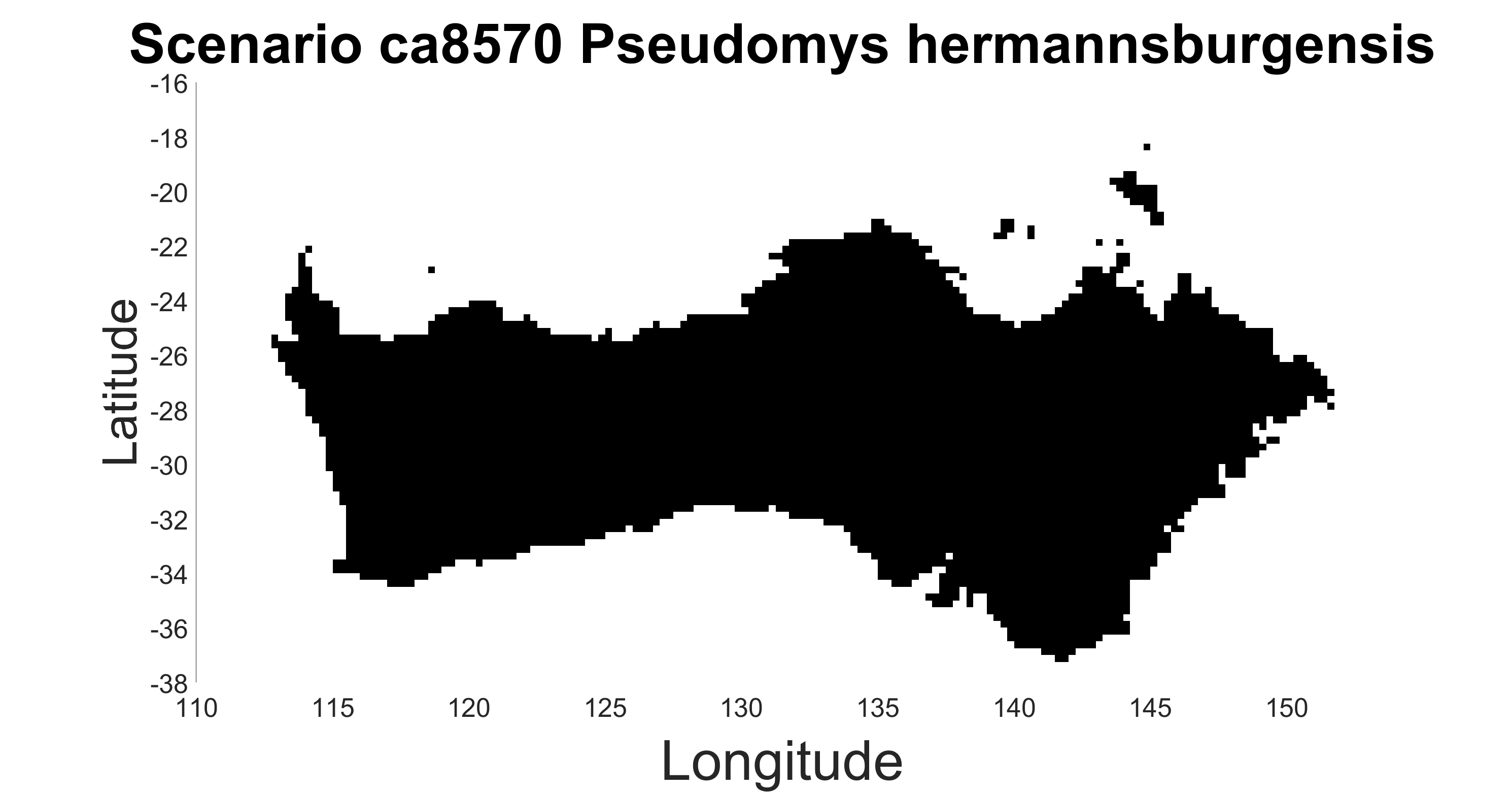}
        \caption{Scenario Range Map }
        \label{fig:weighting:scenario}
        % \caption{Scenario bc2670 Predicted Range Map }
    \end{subfigure}
    \hfill
    \begin{subfigure}{.34\textwidth}
        \centering
        \includegraphics[width=\linewidth]{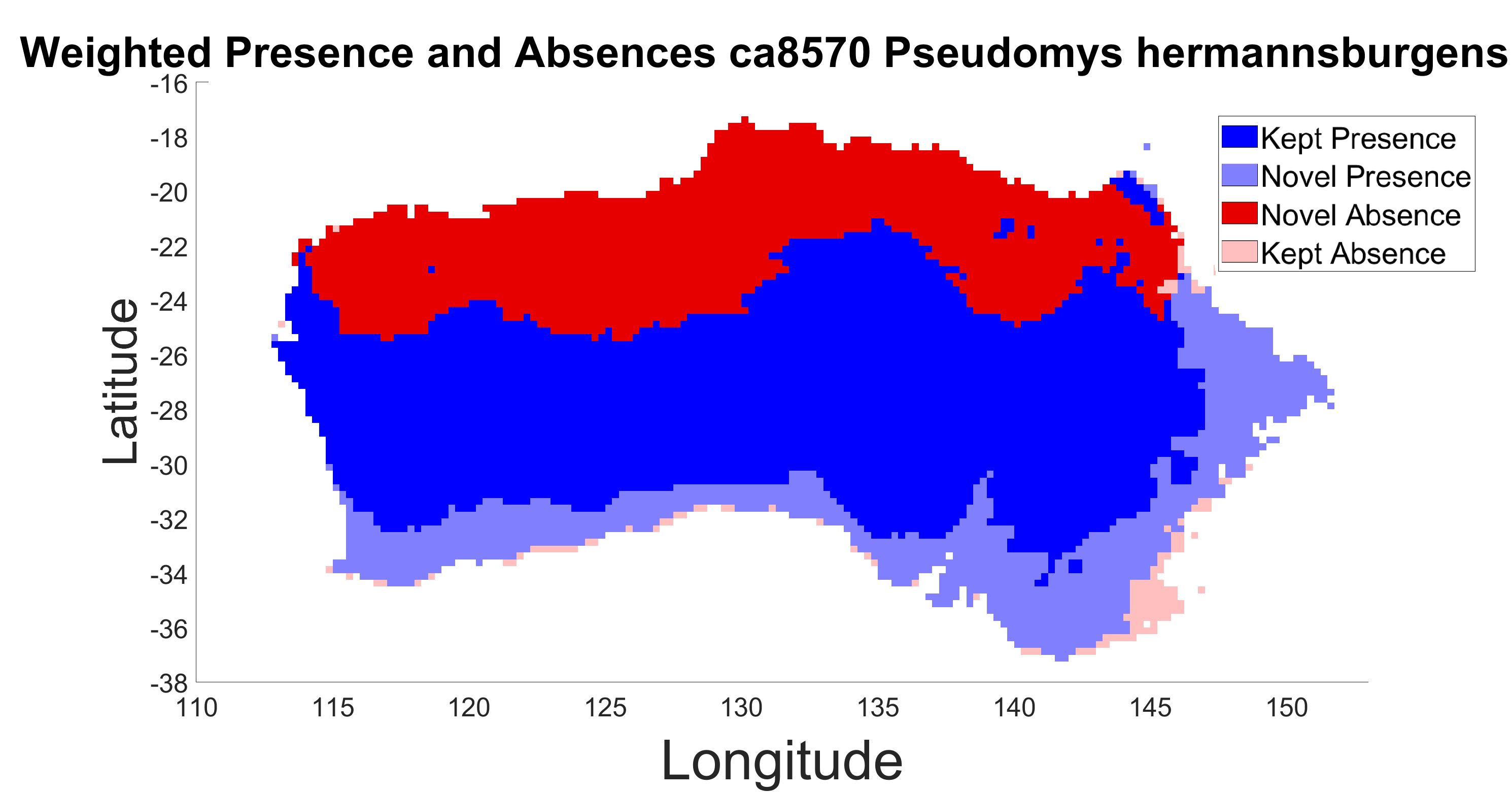}
        \caption{Overlap of Present Day and Scenario}
        \label{fig:weighting:overlap}
    \end{subfigure}
    \caption{Map (a) shows the present day range map for the Australian sandy inland mouse (\textit{Pseudomys hermannsburgensis}). Map (b) is the predicted range map of inland mouse in 2070 under climate scenario bc with RCP 2.6. Map (c) is the corresponding weighted values for each pixel when we intersect (a) and (b): presences are shown in blue, absences in red. Darker color values represent cell weights with greater conservation relevance because they correspond to present cells in (a).}
    \label{fig:weighting} 
\end{figure}

In our weighting scheme, presences are given positive weights, absences negative weights, and we choose $|\an{}| > |\ak{}|$, to emphasize those cells whose ecological suitability for this species is vanishing. These ``lost" cells corresponding to novel absences, $\{(r,c): W_s(r,c) = \an \}$, are particularly important; they represent cells where according to the range map model, the geographic location is no longer suitable given the change in climate. In addition, these cells should be weighted higher to emphasize their importance for conserving a species as they are known to have been occupied at some time point. By contrast the cells corresponding to $\pn{}$ represent regions that are predicted to become suitable, however such predictions do not imply that the locations will be occupied due to uncertainty in dispersal, which is not directly taken into account by the PPPM. Thus, we also make the choice $|\pk{}| > |\pn{}|$. We chose $|\an{}| = |\pk{}| = 1,  |\ak{}| = |\pn{}| = 0.5$ to represent the fact we are more concerned about all cells in the region of present day occurrences. However, we emphasize the flexibility of our method as alternative choices can be made depending on the goals. A visualization of this weighting scheme is shown in Figure \ref{fig:weighting}.

Alternative weightings can be considered for various applications, for instance the choice of $|\pk{}| = |\pn{}| = |\an{}| = |\ak{}| = 1 $ does not make use of the present day map, and does not differentiate cells that underwent a change from the present day map to the predicted scenarios. 

Alternative similarity metrics could have been considered. However, the cosine-similarity has certain advantages over other commonly used measures to quantify the similarities (or distances) between range maps. The Hellinger distance and Kullback-Leibler divergence \citep{hagen2002multi} rely on a probabilistic interpretation, and thus it is difficult to incorporate absences into these measures, however they might be useful when considering differences in range maps describing occurrence probability. The Kappa statistic \citep{hagen2002multi} can be used to measure similarities between categorical maps, however it is not as clear how to weight important cells, such as novel absences. In addition, the Kappa statistic does not directly incorporate the relative frequencies of absences and presences. For instance if two range maps both predict all presences except a single absence cell, these two range maps will have a negative Kappa statistic if their one absence cell differs in location. However, for our purposes we would like to consider such a pair of range maps to be very similar. 
The Wasserstein distance \citep{peyre2019computational} measures the distance a species moves, suggesting it is appropriate in our setting. However, the Wasserstein distance does not model the disappearance of regions, and attempts at computing the Wasserstein distance proved too slow or even non-convergent in all but the range maps with the smallest number of presence cells.

The computation and resultant cosine similarity using these weighted range maps are readily interpretable; when two scenarios agree on the presence or absence of a cell, this cell has a positive contribution to the cosine similarity, whereas the cell has a negative contribution when the two scenarios disagree. For any choice of weights, the resulting similarity is always in the range $[-1,1]$. If the weights are chosen such that $|\an{}| = |\pk{}|, \ |\ak{}| = |\pn{}|$ (as we have assumed), the cosine similarity is $1$ when the two range maps are identical and $-1$ if the two range maps completely disagree on presences. This interpretability allows for a simple method to combine information across species. Furthermore, cosine similarity can also be implemented very efficiently and quickly utilizing matrix multiplication.

\subsubsection{Spectral Clustering} \label{subsubse3:Methodology}
\label{sec:spec}
For each species $m = 1,\dots, 1101$, we compute the pairwise scenario similarity matrix by computing the cosine similarity \eqref{eq:cosSim} on each pair of scenario weighted range maps, ($W^m_s, W^m_{s'}$), where $W^m_s$ is the weighted range map for species $m$ under scenario $s$. That is for each mammal species, we construct the $\bm{s}$ by $\bm{s}$ matrix $S^m$ with entries $S^m(s,s') = CS(W^m_s,W^m_{s'})$, where $\bm{s} = 34$ is the number of scenarios. We cluster scenarios by spectral clustering on the cosine similarity measure \citep{von2007tutorial}.

The properties of spectral clustering are understood from a graph theory perspective \citep{von2007tutorial}. The similarity matrix $S^m$ can be thought of as an undirected graph whose nodes are the scenarios and the edge weight between a pair of scenarios $s$ and $s'$ is given by $S^m(s,s')$. Spectral clustering has best performance on sparse graphs, thus for the dense similarity matrix $S^{m}$, the first step is to sparsify it by taking the $k$-nearest neighbor graph, that is retaining an edge from $s$ to $s'$ only if $s'$ is within the top $k$ neighbors of $s$ (i.e., it is within the top $k$ scenarios of maximal similarity to $s$). When the graph is expressed as a matrix, sparsifying the graph corresponds to setting entries of the matrix to zero. Retaining the top $k$ neighbors leads to a directed graph as this definition of nearest neighbor is not symmetric. Thus we retain the undirected edge from $s$ to $s'$ if either $s'$ is within the top $k$ neighbors of $s$, or vice-versa (Figures \ref{fig:spectralClusteringEx} A to B). The retained edges are weighted by the similarity of their endpoints. We denote by $E$ this matrix of retained weights.

The main computational tool of spectral clustering is the graph Laplacian $L = D - E$, where $D$ is a diagonal matrix of node degree, $D(s,s) = \sum_{s'=1}^{n_s} E(s,s')$ (i.e., the number of scenarios connected to scenario $s$). The graph Laplacian can be thought of as a matrix representation of a graph, with useful mathematical properties as will be discussed shortly. We used the random-walk normalized graph Laplacian, $L_{rw} = D^{-1}L$, as suggested in \cite{von2007tutorial}, although this normalization choice is most significant when the node degrees vary significantly, which is not the case here. The normalization is used to prevent a single node from dominating the resulting spectral clustering when that node has many more connections, i.e., it is a ``central node". However, such ``central nodes" were not found in this application since each scenario is similar to only a few other scenarios (i.e. those with similar RCP and or GCM).

If the graph corresponding to the normalized Laplacian is disconnected (i.e. there is at least one group of nodes that do not have any edges outside that group, Figure \ref{fig:spectralClusteringEx:clean}.C), then there is at least one eigenvector of $L_{rw}$ which is sparse with non-zero entries only on nodes within that group, with an associated eigenvalue of 0. If instead the graph is ``noisier" and fully connected as in our case, the first eigenvectors (corresponding to the smallest eigenvalues) are only approximately sparse; they have large magnitude on groups of nodes that are tightly connected, and small magnitude on nodes outside such groups. Thus these eigenvectors can be used to identify clusters, which is discussed next.

%Both $L$ and $L_{rw}$ have several nice properties, they are positive semi-definite and the multiplicity of their zero eigenvalue is the number of connected components of the graph. For most real-world graphs including the sparsified cosine similarity matrix $E$, the graph is fully connected and thus the number of connected components of this graph is one. When this is the case, the eigenvectors corresponding to the smallest non-zero eigenvalues can be used as an embedding. This can be thought of from a perturbation perspective, if the graph had true clusters of connected components then these eigenvectors would have the same span as vectors representing clustering membership indicators. However in real world graphs there are a few ``noisy" edges between clusters, thus these first few eigenvectors instead are near piecewise on those indicators. 

%Drawing from this insight, the space associated with the eigenvectors of $L_{rw}$, which we denote as columns of a matrix $U$, is used as ``spectral embedding". We use the second and third eigenvectors of $U$ (corresponding to the second and third smallest eigenvalue) as the spectral embedding, that is scenario $s$ is represented by $(U_{s,2},U_{s,3})$.

The spectral embedding of the $\bm{s}$ by $\bm{s}$ matrix $L_{rw}$ is related to the eigendecomposition of $L_{rw}$. The eigenvectors of $L_{rw}$ arranged by decreasing order of the corresponding eigenvalues form the columns of a $\bm{s}$ by $\bm{s}$ matrix $U$. The rows of $U$ represent the coordinates of the corresponding observational unit (in our case observational units are scenarios). This embedding (transformation from a graph, $L_{rw}$, to real valued coordinates, $U$) is meaningful because of the properties of spectral clustering disucssed above.

Illustrations of spectral clustering on both a `clean' (perfectly separable clusters) and `noisy' (one cross edge) examples are shown in Figures \ref{fig:spectralClusteringEx:clean} and \ref{fig:spectralClusteringEx:noisy}. The images illustrate the steps of the spectral clustering described above. The clean example is `cleanly' disconnected: the first four nodes form a cluster which share no edges to the last four nodes (i.e., the symmetrized kNN matrix, Figure \ref{fig:spectralClusteringEx:clean}.B, is a block matrix). The spectral embedding process starting on the clean example can be summarized as follows: it starts with an input similarity matrix (Subfigure A)), where each row (and column) of the matrix represents an observation, and entry (i,j) of the matrix represents the similarity between observations i and j. The next step is to compute the nearest neighbor similarity matrix, which is shown in Subfigure B). Here we have chosen k=3 (the number of nearest neighbors is 3), so we set an entry in the nearest neighbor matrix to 1 if it is within the top 3 values in its row or column, else we set it to 0. The next step is to compute the graph Laplacian $L = D - E$ of the nearest neighbor graph, should in Subfigure C). Here $L$ is the graph Laplacian, $D$ is a diagonal matrix of the node degrees (in this case every node has degree 3) and $E$ is the nearest neighbor graph. This is then normalized to random walk flavor of the graph Laplacian, $L_{rw} = D^{-1}L$. The final step is to compute the spectral embedding from the graph Laplacian, which is done by taking the smallest two eigenvectors of the random walk graph Laplacian, shown in Figure D). The two columns serves as the two coordinates: the first column is the x-coordinate in subplot D), the second column the y-coordinate. We see that the spectral embedding in Figure D) faithfully represents the original graph, there are still two clusters of four nodes each, but now represented in a two dimensional plane rather than the original similarity matrix.

The noisy example has a single cross edge entry of the matrix (i.e., there is one pair of scenarios from the original separate groups that are now considered neighbors, Figure \ref{fig:spectralClusteringEx:noisy}.B) that affects the final spectral embedding; the first and fifth point are closer together (Figure \ref{fig:spectralClusteringEx:noisy}.D). Our case of 34 scenarios is more complex than this illustrated example of only 8 points, however the pattern that groups of scenarios which are all similar to each other will be near each other in the embedded space remains true.

\begin{figure}
    \begin{subfigure}{1.0\textwidth}
        \centering
        \includegraphics[width=.9\linewidth]{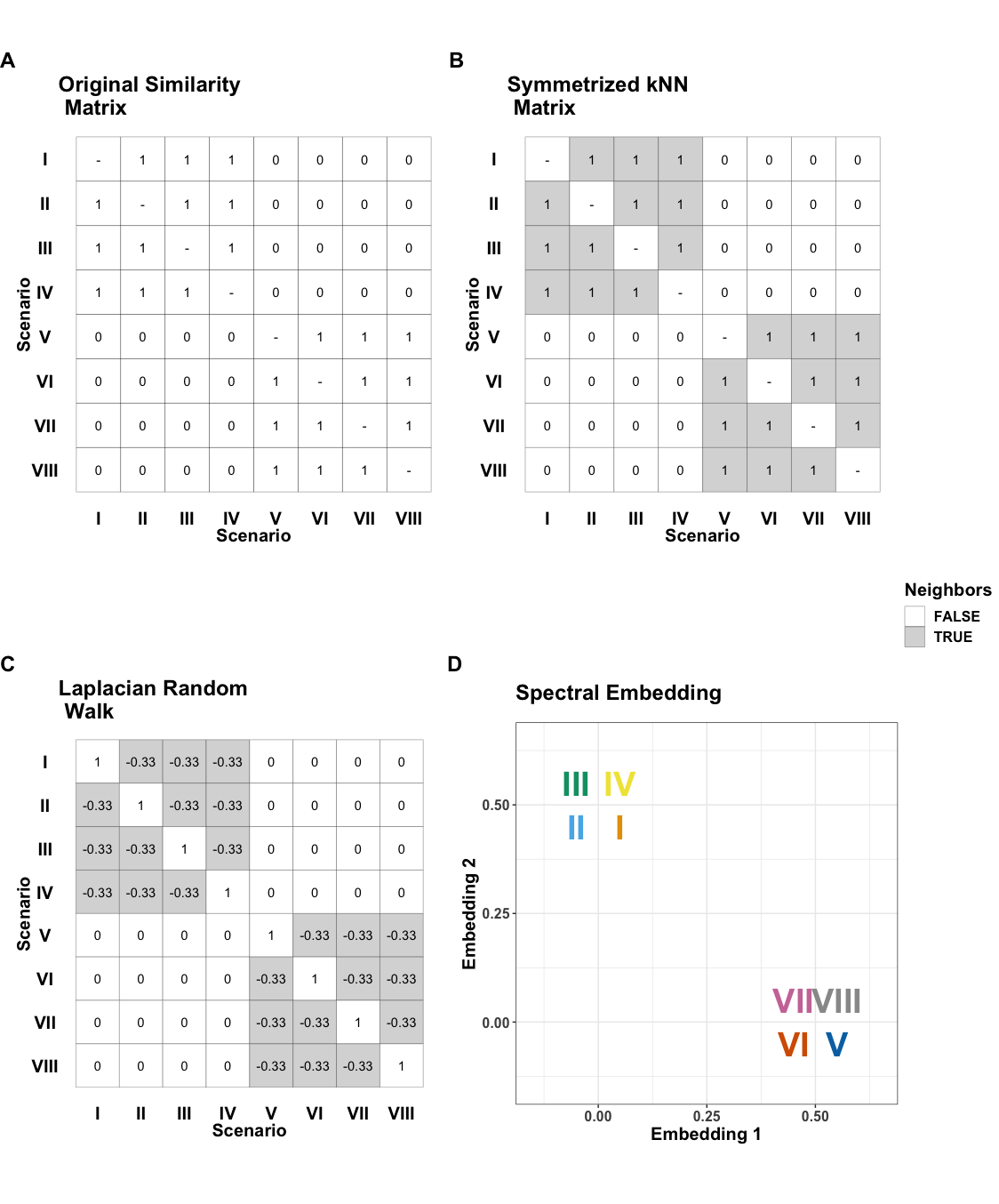}
        \caption{Spectral Clustering Clean Example: the two groups are separable in that there are no shared members. These four plots show the steps of spectral embedding: Subfigure A) is the example input similarity matrix, where each row (or column) represents an observation. Subfigure B) shows the nearest neighbor step with the choice of k = 3, Subfigure C) shows the graph Laplacian (the random walk version of the Laplacian) of the nearest neighbor graph, and D) shows the final spectral embedding.  We see that in this ideal example the spectral embedding perfectly separates the two groups of columns (the first four and second four observations are in separate clusters). The results on this clean example can be compared to a noisier example in Figure \ref{fig:spectralClusteringEx:noisy}.}
        \label{fig:spectralClusteringEx:clean}
    \end{subfigure}
\end{figure}

\begin{figure}
    \ContinuedFloat
    \begin{subfigure}{1.0\textwidth}
        \centering
        \includegraphics[width=.9\linewidth]{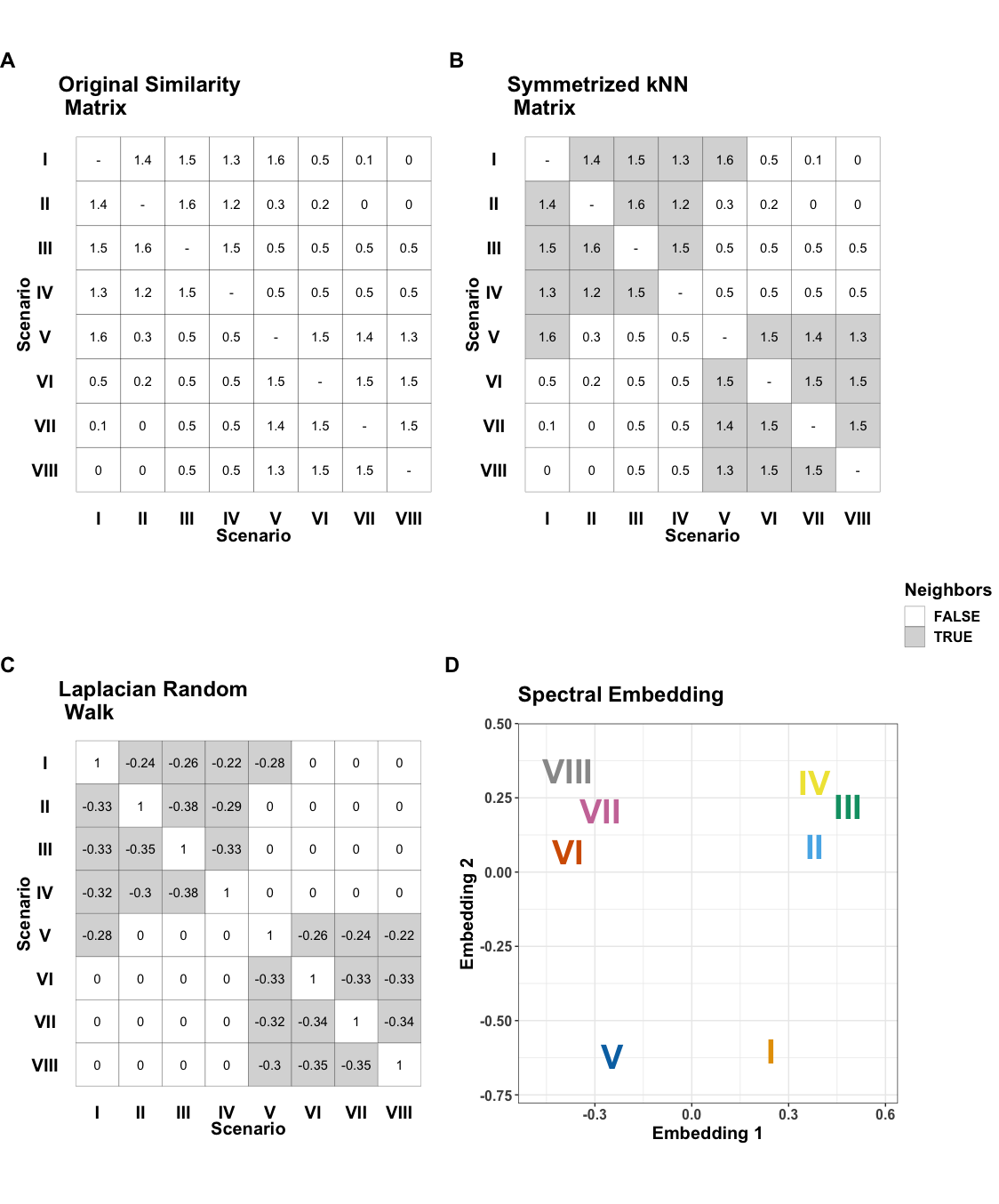}
        \caption{Spectral Clustering Noisy Example: the two groups are no longer separable in that there is one pair of members which are linked across groups, I and V. This linkage is faithfully represented in the spectral embedding by placing groups I and V closer together and separate from the other examples.}
        \label{fig:spectralClusteringEx:noisy}
    \end{subfigure}
    \caption{Step-by-step illustrations of spectral clustering for a) a 'clean' example and b) a 'noisy' example. The noisy example perturbs the clean example to have one pair of neighbors across the original separate groups. For both examples, the original similarity matrix in Subfigure A) defines the neighbors in the symmetrized kNN matrix with k = 3 in Subfigure B). The Laplacian random walk matrix in Subfigure C) is calculated from B) and the two eigenvectors of C) corresponding to the smallest non-zero eigenvalues correspond to the coordinates of the scenarios in Subfigure D). Note in D) there is some jittering for visualization purposes. The perturbation of the similarity matrix results in scenarios I and V being plotted more closely in the noisy example.}
    \label{fig:spectralClusteringEx}
\end{figure}

Once observations (in our case future climate scenarios) are transformed to real coordinates via spectral embedding as discussed above, a simple clustering algorithm can be applied such as k-means. However we chose the single-linkage clustering algorithm over k-means in this embedded space as it performs well according to the Davies-Bouldin criterion. The Davies-Bouldin criterion is a common clustering criterion quantifying the goodness of separation between clusters \citep{davies1979cluster}. The Davies-Bouldin criterion is interpreted as a ratio of the average intra-cluster distance and inter-cluster distance; a good clustering will have points that are close together within cluster (small intra-cluster distance), and points that are far apart in different clusters (large inter-cluster distance).

Spectral clustering can be performed for a single species using only $S^m$. Alternatively, one way to combine information across species is to capture group wide trends in scenarios. We cluster similarity matrices based on two aggregations. First, we average the similarity matrices across all $\bm{m} = 1101$ species, as in $S= (1/\bm{m})\sum_{m=1}^{\bm{m}} S^m$. Second, we average over a subset of species whose fraction of area lost is among the highest $10\%$. We average the similarity matrices across species in order to incorporate information from each species into a quantitative clustering approach, obtaining a broader understanding of the overarching patterns between species range maps and climate scenarios.

\subsubsection{Climate Based Scenario Clustering}
\label{se:climatebased}
    We contrast the clustering based on range maps with a clustering based only on predicted climate variables. This comparison will illustrate the importance of incorporating ecological information by contrasting clusters based only on climate to those using climate to first estimate species ranges before clustering. 
    We discuss the process of clustering using only the five climate variables that were used to predict the range maps (see Section \ref{se:Methods} for a description of the predictions). Each of these five globally distributed variables is predicted across the 34 scenarios. In order to directly compare climate-based clustering to our ecologically based clustering, we utilized the climate variables to perform a clustering in a similar fashion to the ecologically based clustering. However, for continuous data, the cosine similarity is not appropriate, as two maps that are shifted versions of each other would be considered very similar to each other. For instance, if one scenario predicted two degrees Celsius warmer everywhere than another scenario, the cosine similarity between these two maps would be very high, which is not desirable as these two maps represent significantly different predictions. Instead, we use the $L_2$ distance between maps, which will effectively use both the difference between the means of maps and differences in the spatial variation. To incorporate all five climate variables, each variable was normalized before applying the $L_2$ distance. We denote $T_s^f(r,c)$ as the value of variable $f$ according to scenario $s$ at location $(r,c)$. The variable scaled $L_2$ distance between a pair of scenarios $s$ and $s'$ using all five variables is given by
    $$ H(s,s') = \sum_{f = 1}^{5} \sum_{r=1}^{\bm{r}}\sum_{c=1}^{\bm{c}} [(T_s^f(r,c) - T_{s'}^f(r,c))/\sigma_f]^2,$$ 
    where $\sigma_f$ is the standard deviation of variable $f$ measured across all locations and scenarios, that is we calculate the mean of each variable across all locations and scenarios, 
    \begin{equation*}
    \sigma_f^2 = \sum_{s = 1}^{\bm{s}} \sum_{r=1}^{\bm{r}}\sum_{c=1}^{\bm{c}} [T_s^f(r,c) - \mu_f]^2
    \hspace{0.2cm} \text{ with } \hspace{0.2cm}
    \mu_f = (\bm{s}\cdot \bm{r}\cdot \bm{c})^{-1} \sum_{s = 1}^{\bm{s}} \sum_{r=1}^{\bm{r}}\sum_{c=1}^{\bm{c}} T_s^f(r,c).
    \end{equation*}
    
    We create a similarity matrix, $S$, from the distances by using the monotonically negative transformation $S(s,s') := 1/H(s,s')$. Spectral clustering can be performed on this climate-informed matrix S which can be compared to the ecologically driven spectral clustering.
    
\subsection{Rand Index}
In order to quantitatively compare cluster assignments, we used the adjusted Rand index \citep{hubert1985comparing}. The adjusted Rand index measures how similar two cluster assignments are to each other. We make use of the adjusted Rand index by measuring the similarity of the clustering results to partition by either the representative concentration pathway (RCP) or global climate model (GCM), in order to quantitatively answer whether RCP or GCM differences is the driving factor behind the clustering results. 

% Poorly chosen variable letters, I am runing out of variable names, not sure what to choose
Given two cluster assignments $X = \lbrace X_1,\ldots,X_n \rbrace $ and $Y = \lbrace Y_1 ,\ldots, Y_m \rbrace$, where each $X_i$ and $Y_j$ are clusters (sets of scenarios). Define the cardinality of the set of overlaps between two clusters as $o_{ij} := | X_i \cap Y_j |$. Also define $o_{i\cdot}:= \sum_{j=1}^m o_{ij}$ and $o_{\cdot j}:= \sum_{i=1}^n o_{ij}$. The adjusted Rand index is given by

$$ \text{ Adjusted Rand Index } = \frac{\sum_{i=1}^n\sum_{j=1}^m \binom{o_{ij}}{2} - [\sum_{i=1}^n \binom{o_{i\cdot}}{2}\sum_{j=1}^m \binom{o_{\cdot j}}{2}]/\binom{n}{2}}{\frac{1}{2}[\sum_{i=1}^n\binom{o_{i\cdot}}{2} + \sum_{j=1}^m \binom{o_{\cdot j}}{2}] - [\sum_{i=1}^n\binom{o_{i\cdot}}{2}\sum_{j=1}^m \binom{o_{\cdot j}}{2}]/\binom{n}{2}}, $$
where $\binom{n}{2}$ denotes the binomial coefficient and is calculated as $\binom{n}{2} = \frac{n(n-1)}{2}$.

The adjusted Rand index measures how similar two cluster assignments are to each other compared to random assignments. Random assignments have an expected adjusted rand index of 0.

\section{Results and Discussion} \label{se:resulstanddis}

\subsection{Global Diversity Loss} \label{sec:glob}
\begin{figure}
    \begin{subfigure}{.5\textwidth}
        \centering
        \includegraphics[width=.9\linewidth]{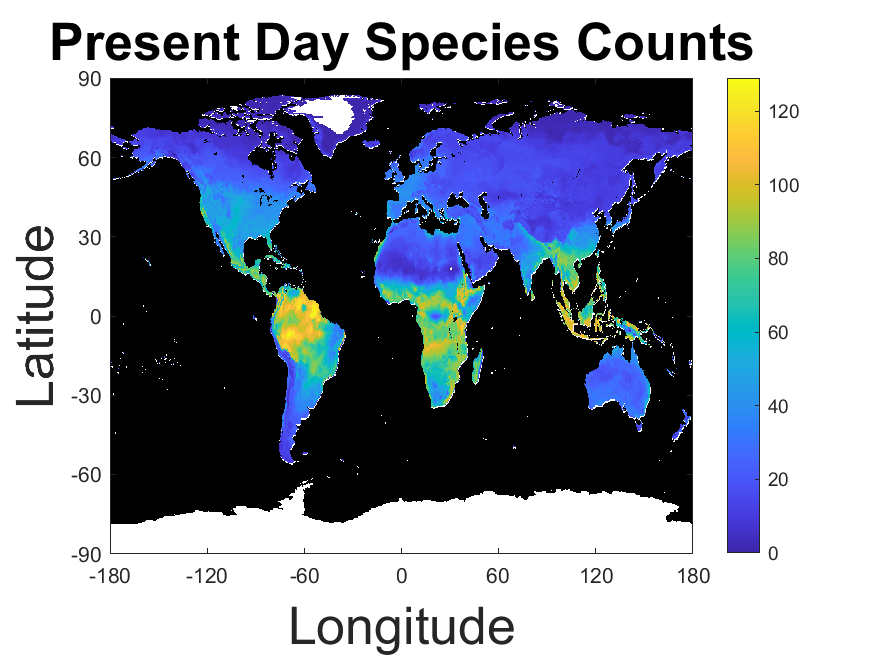}
        \caption{Present Day Mammal Richness}
        \label{fig:globalRisk:present}
    \end{subfigure}
    \hfill
    \begin{subfigure}{.5\textwidth}
        \centering
        \includegraphics[width=.9\linewidth]{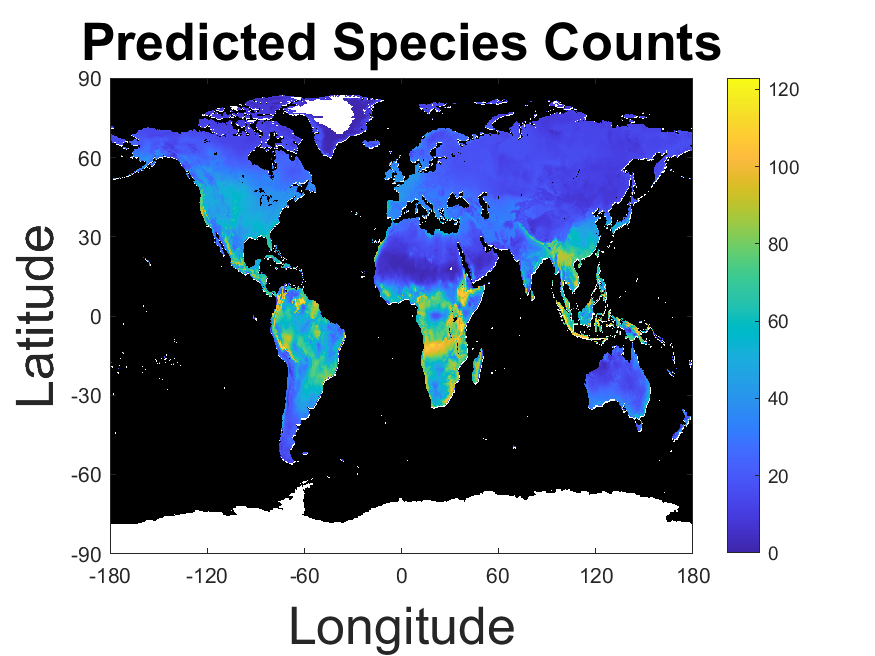}
        \caption{Predicted Mammal Richness}
        \label{fig:globalRisk:predicted}
    \end{subfigure}
    \hfill
    \begin{subfigure}{.5\textwidth}
        \centering
        \includegraphics[width=.9\linewidth]{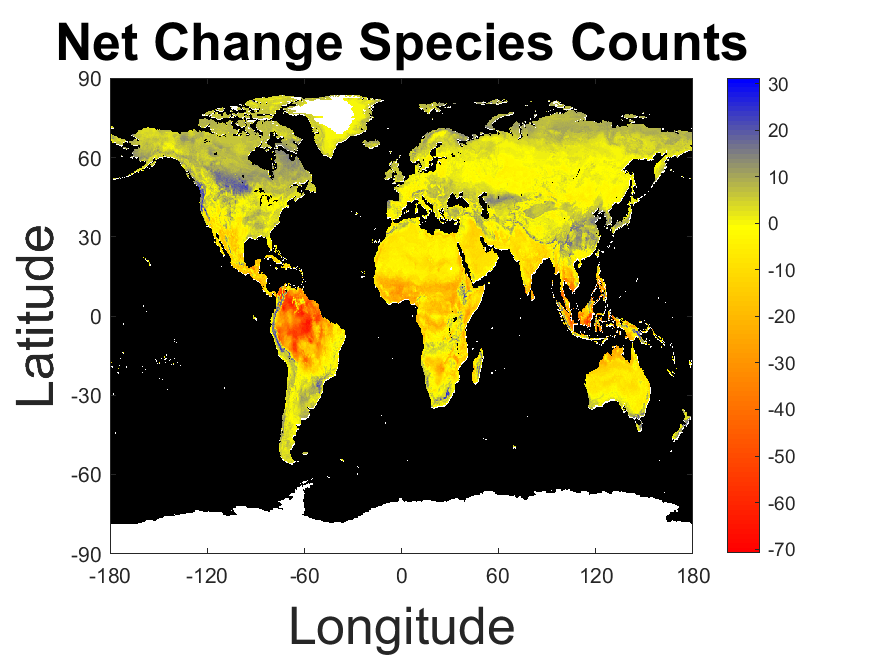}
        \caption{Net Change}
        \label{fig:globalRisk:net}
    \end{subfigure}
    \hfill
    \begin{subfigure}{.5\textwidth}
        \centering
        \includegraphics[width=.9\linewidth]{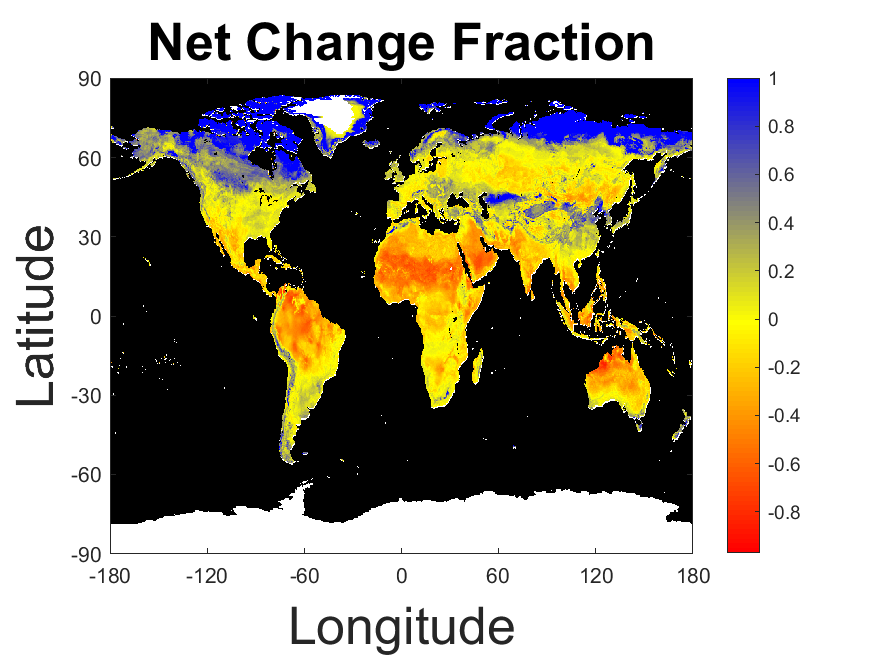}
        \caption{Fraction Change}
        \label{fig:globalRisk:fraction}
    \end{subfigure}
    \caption{ Visualization of the mammal richness in the dataset over space. White cells correspond to locations with no predicted presences. \ref{fig:globalRisk:present} : Present day mammal richness. \ref{fig:globalRisk:predicted}: Predicted Counts averaged over all scenarios. \ref{fig:globalRisk:net}: Net Change, that is, the difference between the Predicted and the Present count. \ref{fig:globalRisk:fraction}: Fraction Change, that is, the fraction between the Net Change and the Present count. Our findings are consistent with \cite{chen2011rapid} which finds that species are moving poleward and towards higher elevations, there is loss around the equator and some increase in diversity towards the northern pole.}
    \label{fig:globalRisk}
\end{figure}

One can get an overall sense of the changes predicted in the range maps from Figure \ref{fig:globalRisk}, which shows how the diversity of mammals is spread spatially over the earth. We see from both Figures \ref{fig:globalRisk:net} and \ref{fig:globalRisk:fraction} that significant potential diversity losses are predicted around the equator in South America and Africa, whereas there is some potential diversity increases further north, consistent with previous findings \citep{chen2011rapid}. We note that these `losses' do not necessarily imply a local extinction but rather that a species is exposed to climate beyond its current realized niche. Similarly species `gains' do not necessarily imply that new species will occur in a given cell, but rather that the cell is newly suitable for a species climatically. A species with newly suitable locations may still not occur there due to dispersal limitation, biotic interactions, etc.

\subsection{Potential Range Shifts} \label{subse1:resulstanddis}
\begin{figure}[ht]
\centering
\includegraphics[width=0.47\textwidth]{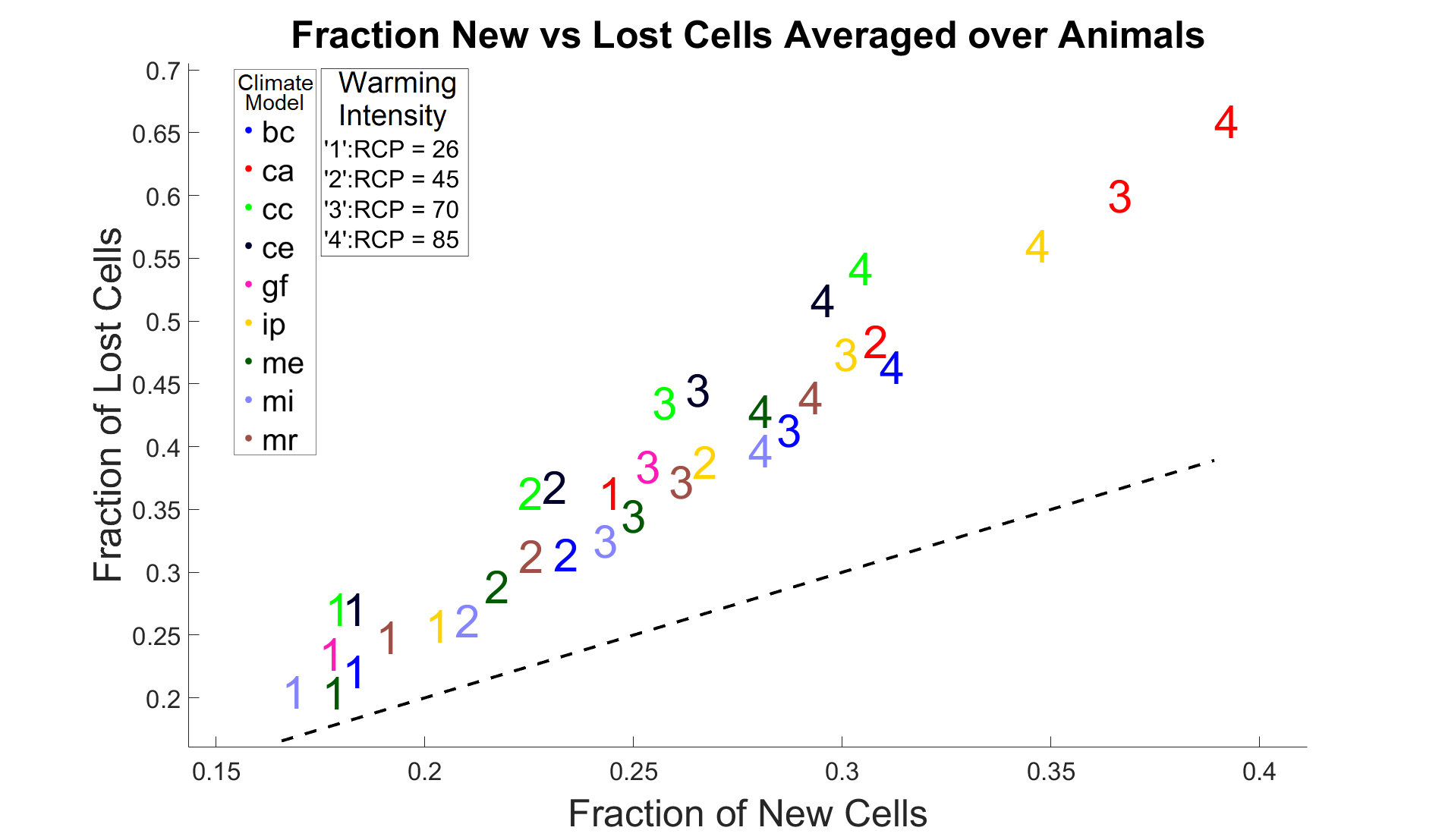}

\caption{Shows that predicted range maps tend to be shifted, that is, lost and new cells grow together, with losses being larger than gains, as the scenarios are above the 45 degree dashed black line.}
\label{fig:scenStats}
\end{figure}

By comparing the fraction of cells corresponding to $\an{}$ and $\pn{}$, we get a sense of potential range map shifts compared to  present day range maps. This is shown in Figure \ref{fig:scenStats}, which shows potential range map shifts, as opposed to range map expansions or contractions; the number of new presences and new absences grow together, which would occur as range map shifts, instead of say absences growing as presences shrink, which would be indicative of an overall range size decrease. 
    
However, although novel absences and novel presences tend to occur together as shown in Figure \ref{fig:scenStats}, on average there are more novel absences (average 38\% $\pm$ 28\% s.d. of current range map) than novel presences (average 25\% $\pm$ 26\% s.d.), which implies that the species ranges are both shifting and decreasing in size (average 13\% net loss, $\pm$ 31\% s.d.) due to the changing climate. 

\subsection{Clustering Plots} \label{subse2:resulstanddis}
\begin{figure}
    \begin{subfigure}{.5\textwidth}
        \centering
        \includegraphics[width=.9\linewidth]{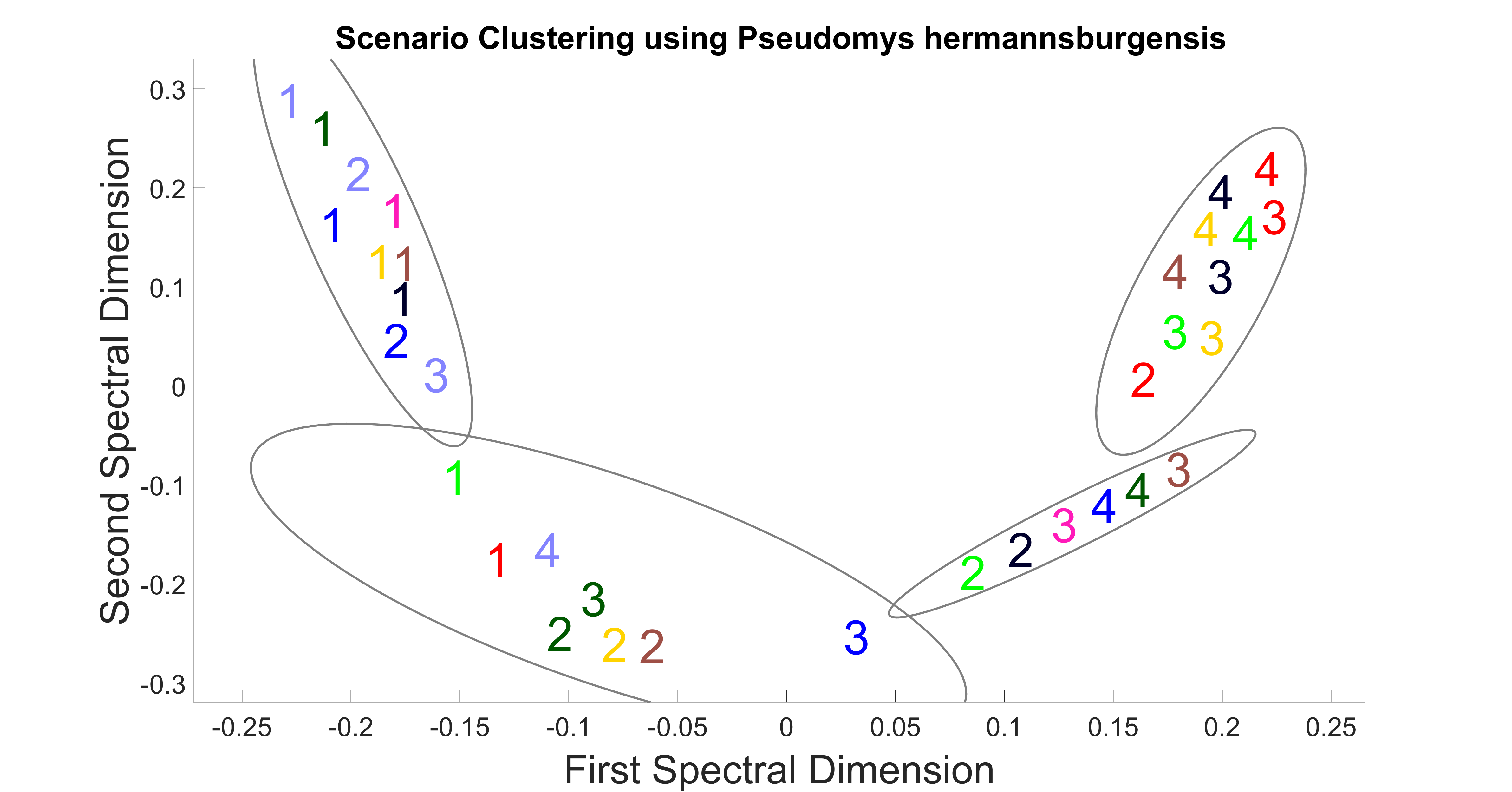}
        \caption{\emph{Pseudomys hermannsburgens} (mouse)}
        \label{fig:mammalClustering:cheetah}
    \end{subfigure}
    \hfill
    \begin{subfigure}{.5\textwidth}
        \centering
        \includegraphics[width=.9\linewidth]{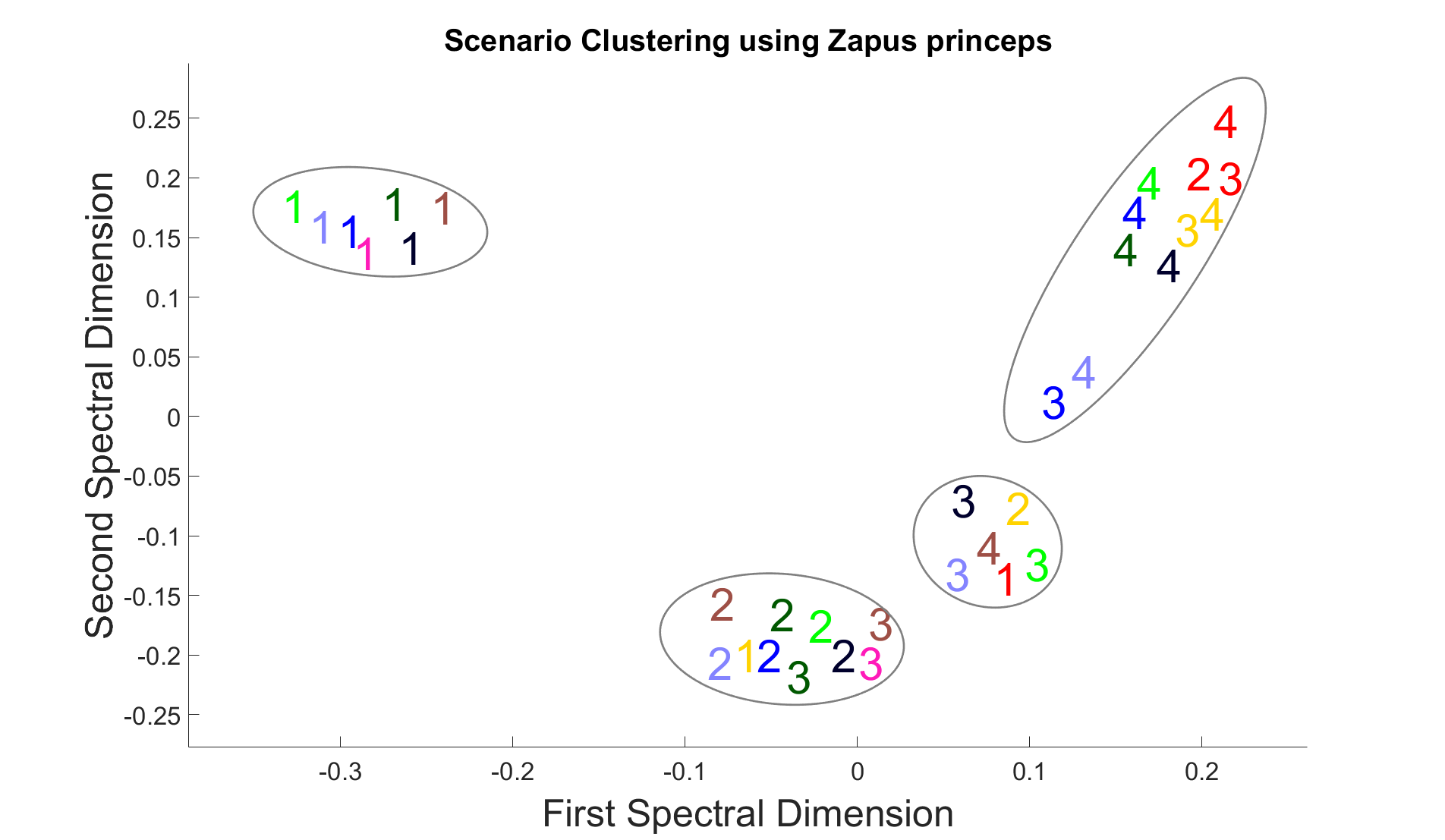}
        \caption{\emph{Zapus princeps}}
        \label{fig:mammalClustering:princeps}
    \end{subfigure}
    \hfill
    \begin{subfigure}{.5\textwidth}
        \centering
        \includegraphics[width=.9\linewidth]{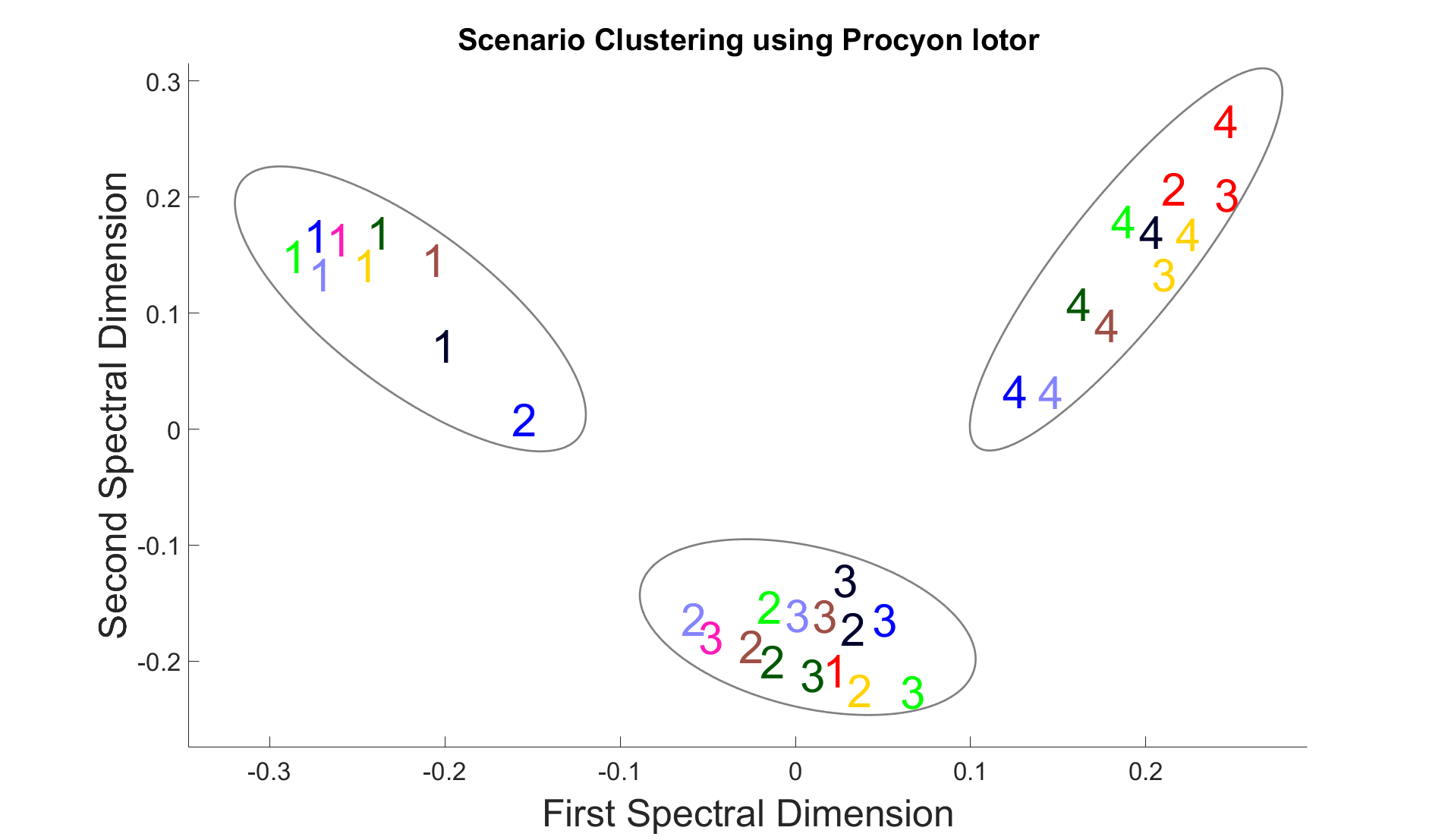}
        \caption{\emph{Procyon lotor}}
        \label{fig:mammalClustering:procyon}
    \end{subfigure}
    \hfill
    \begin{subfigure}{.5\textwidth}
        \centering
        \includegraphics[width=.9\linewidth]{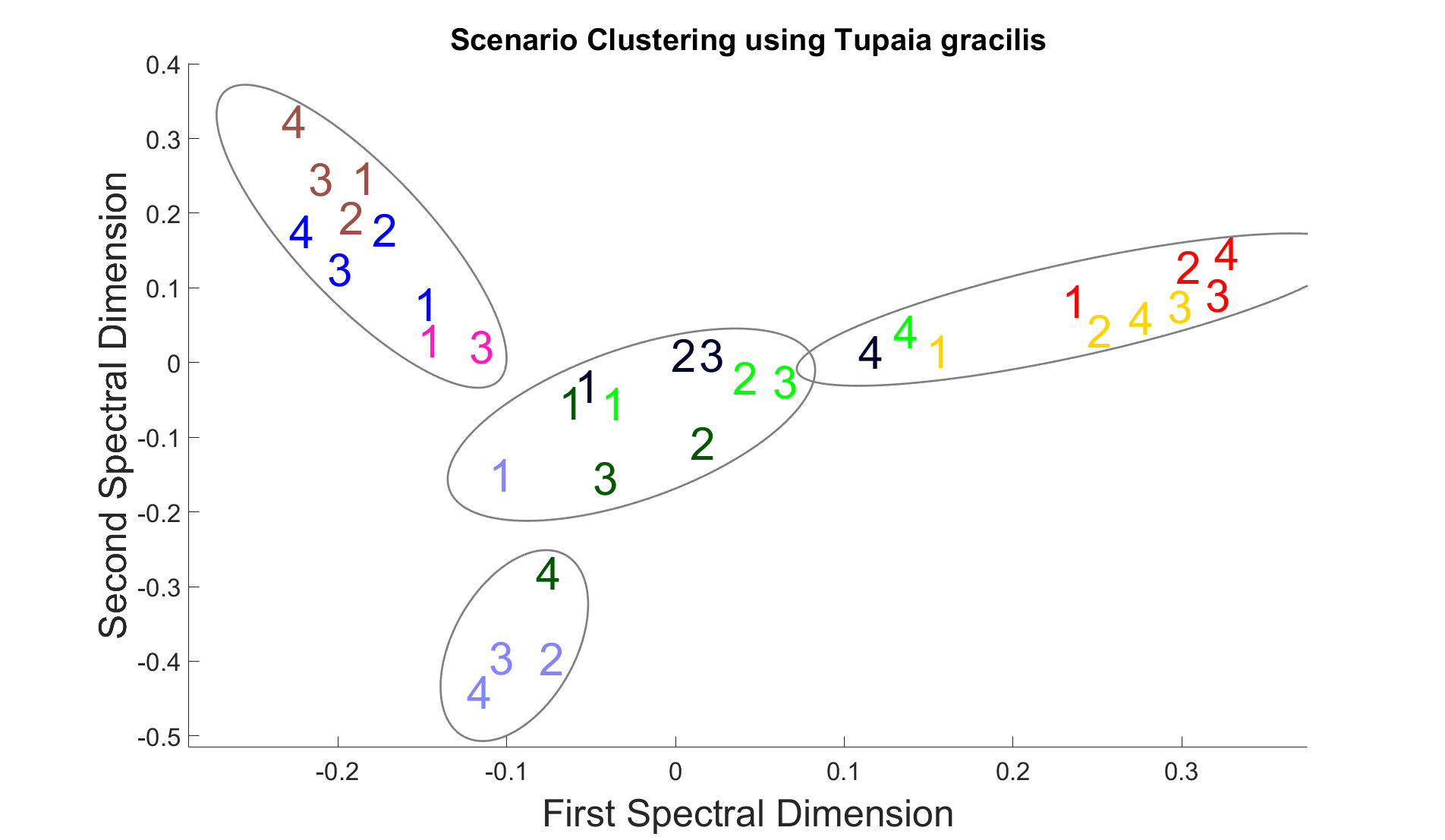}
        \caption{\emph{Tupaia gracilis} (treeshew)}
        \label{fig:mammalClustering:gracilis}
    \end{subfigure}
    \caption{Spectral clustering of scenarios using individual species. We see mainly grouping by RCP for the first three (Figures \ref{fig:mammalClustering:cheetah}--\ref{fig:mammalClustering:procyon}), but a starkly different GCM-driven clustering for the treeshew (Figure \ref{fig:mammalClustering:gracilis}). This variety was found among species, most species clusterings are strongly driven by RCP, but some are more reflective of GCM. These differences between species further demonstrates that ecological information is important to interpret the differences between scenarios.}
    \label{fig:mammalClustering} 
\end{figure}

% This needs more explanation: need to explain where these 30 and 70 percent come from, requiring explanation of RAND index
The spectral clustering results using the species-specific similarity matrix, $S^m$, is shown in Figure \ref{fig:mammalClustering} for four species, illustrating the main types of patterns observed across all 1101 species. We found an interesting mixture of RCP and GCM dependence.  RCP is a major driving factor of cluster composition; in most of the species-specific clusterings, the far left (``optimistic") cluster contains mainly scenarios with low RCP, and the far right (``extreme") cluster only scenarios with high RCP (Figures \ref{fig:mammalClustering:cheetah}--\ref{fig:mammalClustering:procyon}). However, the clustering for the slender treeshew (\emph{Tupaia gracilis}; Figure \ref{fig:mammalClustering:gracilis}) predicted ranges is driven mainly by GCM. This clustering mainly by GCM was found in many species (30\% species' cluster results had a higher Rand index with a GCM clustering than an RCP clustering), probably because these species niches had relatively weak dependence on mean annual temperature compared to other climate variabels. However, the most common trend is RCP dependence (70\% of species).  This variability among species is further evidence supporting the importance of the climate-ecology relationship; the most important difference between climate models (GCM or RCP) varies depending on the individual species.

% TO-DO: Explain looked for spatial patterns/patterns of size, did not find anything significant. Future research to find significant relationships

\begin{figure}
    \begin{subfigure}{.5\textwidth}
        \centering
        \includegraphics[width=.9\linewidth]{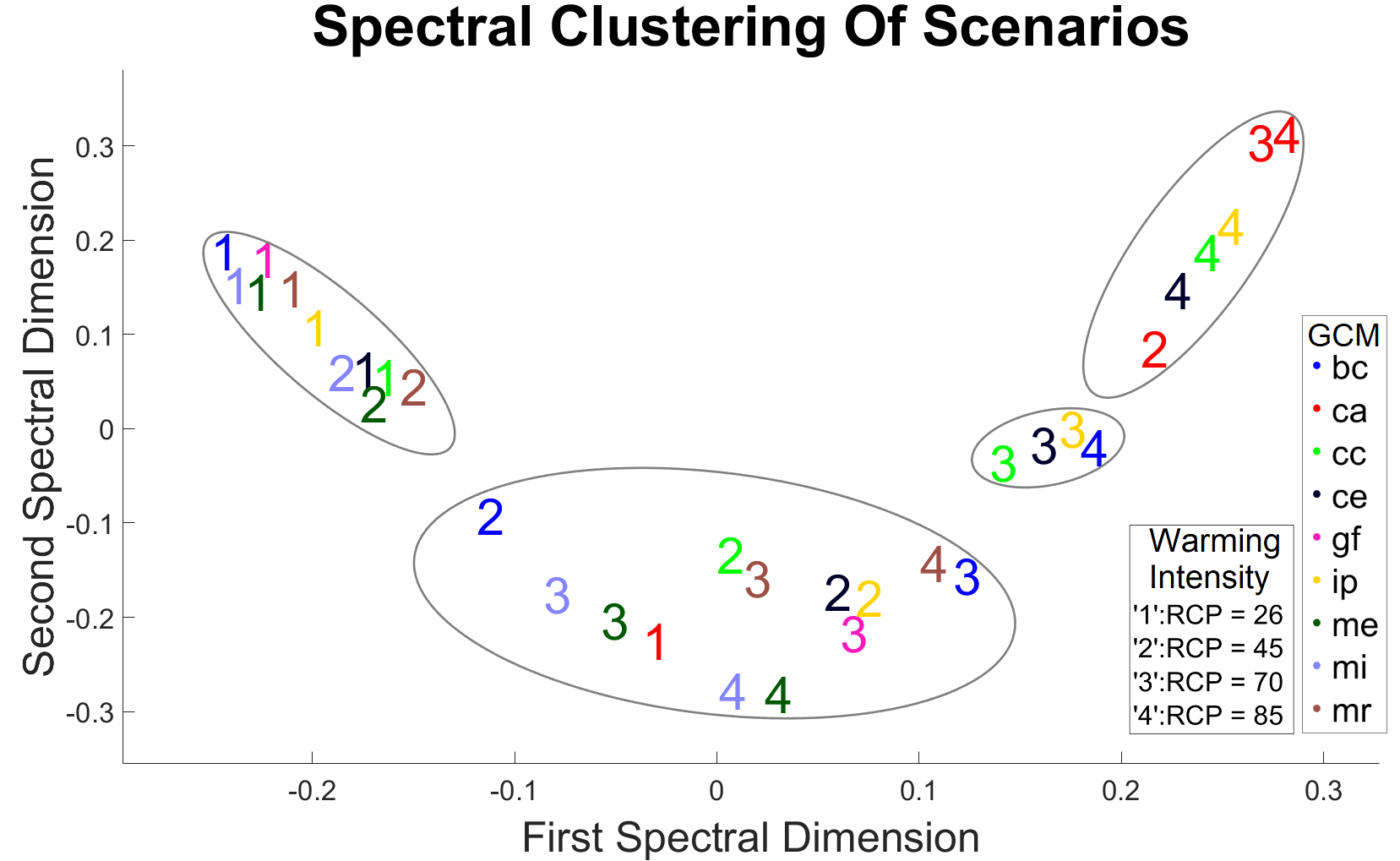}
        \caption{Predicted Range Map Based}
        \label{fig:climateClustering:rangeMap}
    \end{subfigure}
    \begin{subfigure}{.5\textwidth}
        \centering
        \includegraphics[width=.9\linewidth]{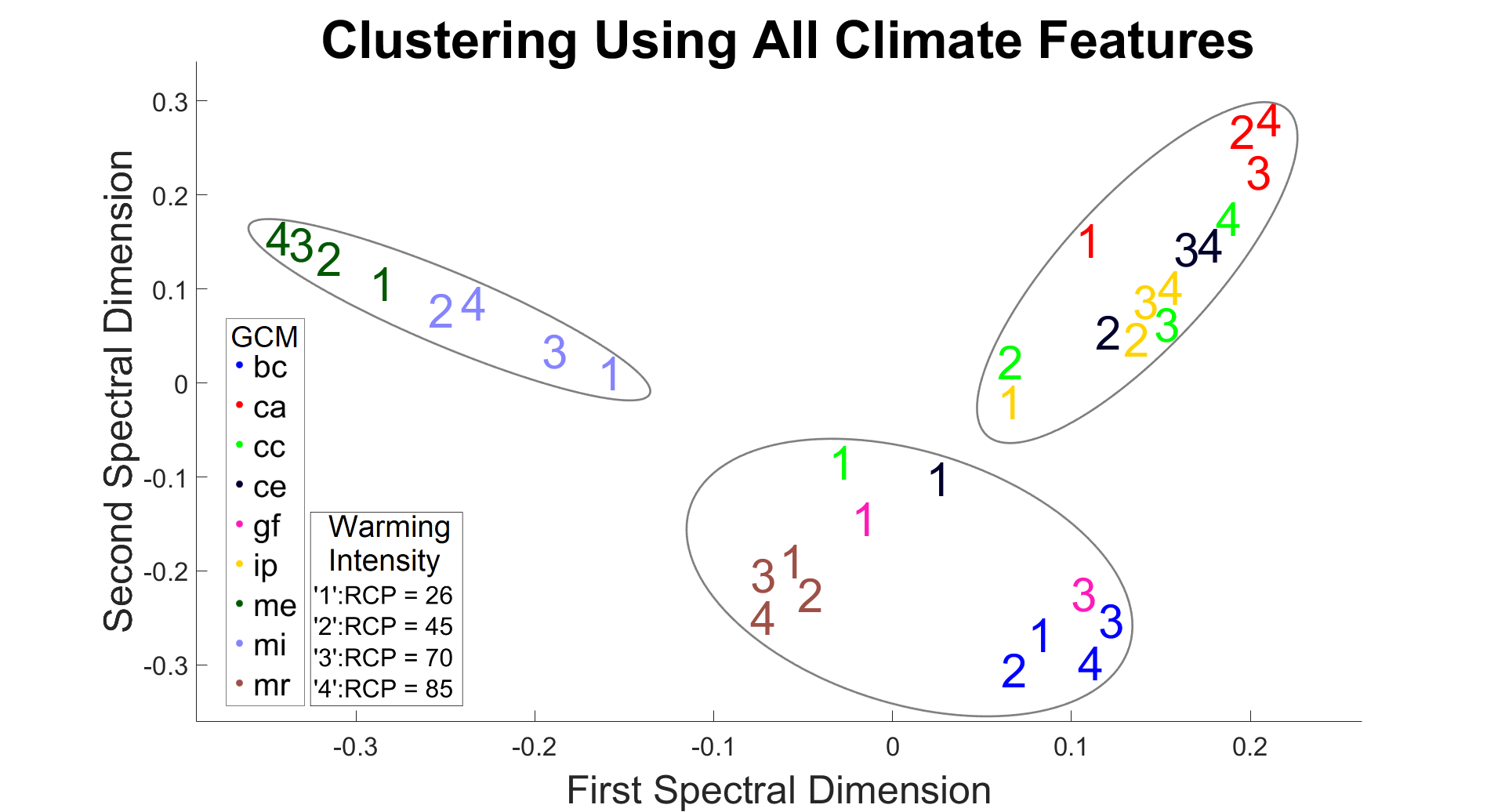}
        \caption{Climate Variable Based}
        \label{fig:climateClustering:climate}
    \end{subfigure}
    \caption{(a): Scenario embedding and clustering for similarity matrix averaged over all 1101 mammals, $S$. The clusters are mainly driven by RCP. However, there is still some relationships among the GCM, for instance the red `ca' GCM predicts more extreme outcomes (right most cluster), and at each level of RCP the green, black, and yellow (`cc',`ce',`ip') climate models are nearby in the spectral embedding space. (b): In the climate-based clustering, the clusters are mainly determined by GCM. This clustering is significantly different from the ecological based one (a), which demonstrates the importance of incorporating ecological information. }
    \label{fig:climateClustering} 
\end{figure}

The spectral cluster results from using the similarity matrix averaged over all 1101 mammals considered, $S$, is shown in Figure \ref{fig:climateClustering:rangeMap}. We see in the bottom cluster of Figure \ref{fig:climateClustering:rangeMap} a mix of RCP and GCM. Similar to the individual species results, this suggests that RCP alone does not account for the variation, and the GCM is also important. 

\begin{figure}[ht]
\centering
\includegraphics[width=1.0\textwidth]{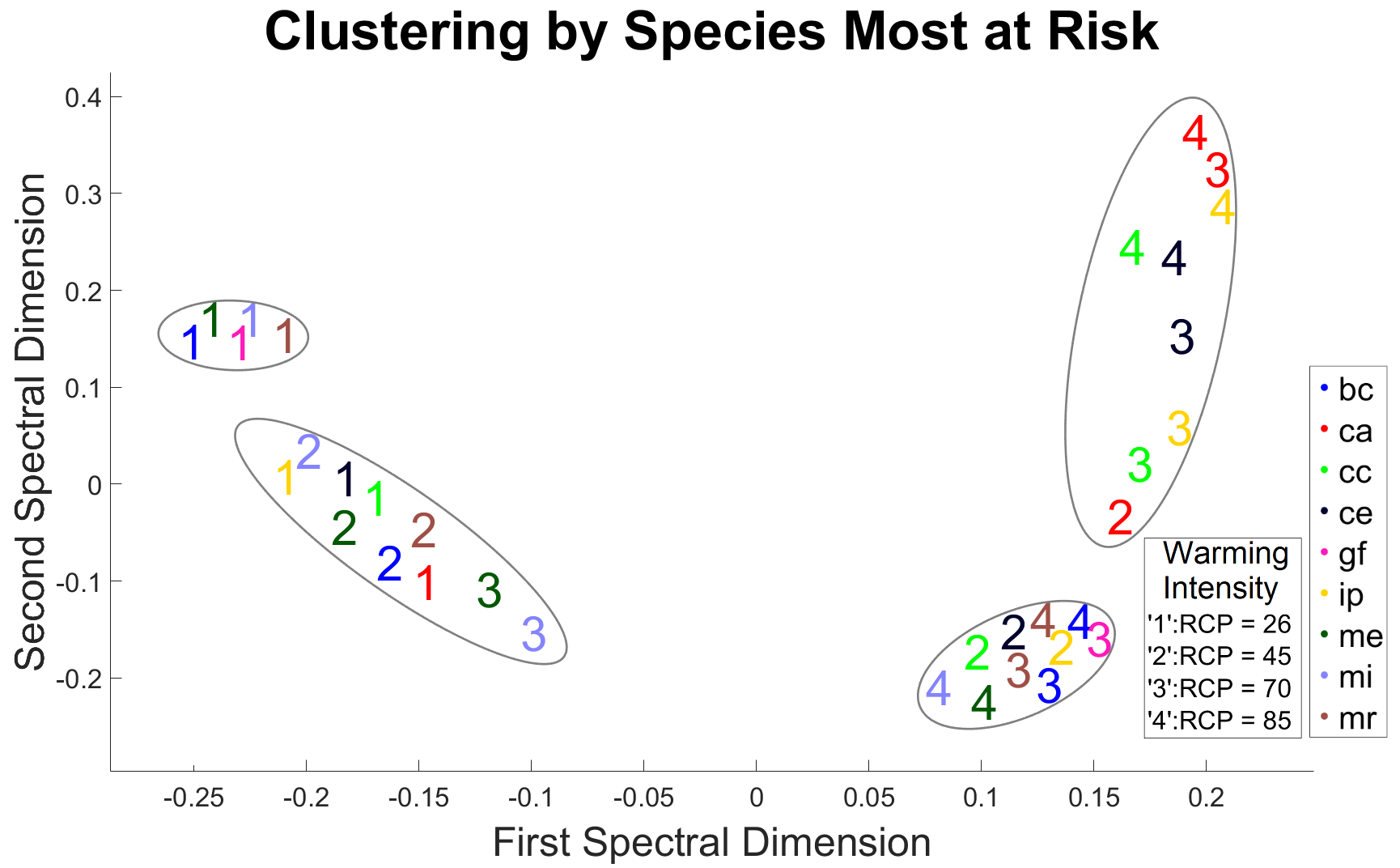}
\caption{Scenario embedding and clustering using only species whose fraction of area lost is among the highest $10\%$. Clustering based only on these species most at risks puts an even higher emphasis on RCP than the clustering in Figure \ref{fig:climateClustering:rangeMap}. This further demonstrates the importance of accounting for ecological information, different subsets of ecological populations emphasis RCP even more strongly.}
\label{fig:Q90}
\end{figure}

We also performed clustering for only the species most at risk, defined by those species whose fraction of area lost is among the highest $10\%$. This loss in area can be used to approximate a loss in population abundance using the techniques in \cite{he2012area} and \cite{che2016species}. Performing spectral clustering by using the average of similarity matrices of this subset of species at risk is shown in Figure \ref{fig:Q90}. This clustering puts a strong emphasis on RCP, that is, the species most at risk are more sensitive to RCP.

\subsection{Cluster Quality Analysis} \label{subse3:resulstanddis}
 One way to qualitatively measure the performance of the clustering is to look at the spatial overlap of presences for each scenario within the clusters, $d$. That is, we define the frequency map $F^m_d$ for species $m$ in cluster $d$ based on the range maps as $F^m_d(r,c):=\sum_{s \in d} B^m_s(r,c)$. This frequency analysis provides qualitative evidence for the quality of the cluster results. A clustering of high quality should have similar members in the same cluster, and dissimilar members in different clusters. In this spatial application, similar members should have high overlap of presence cells.

\begin{figure}
    \centering
    \begin{subfigure}{.45\textwidth}
        \centering
        \includegraphics[width=\linewidth]{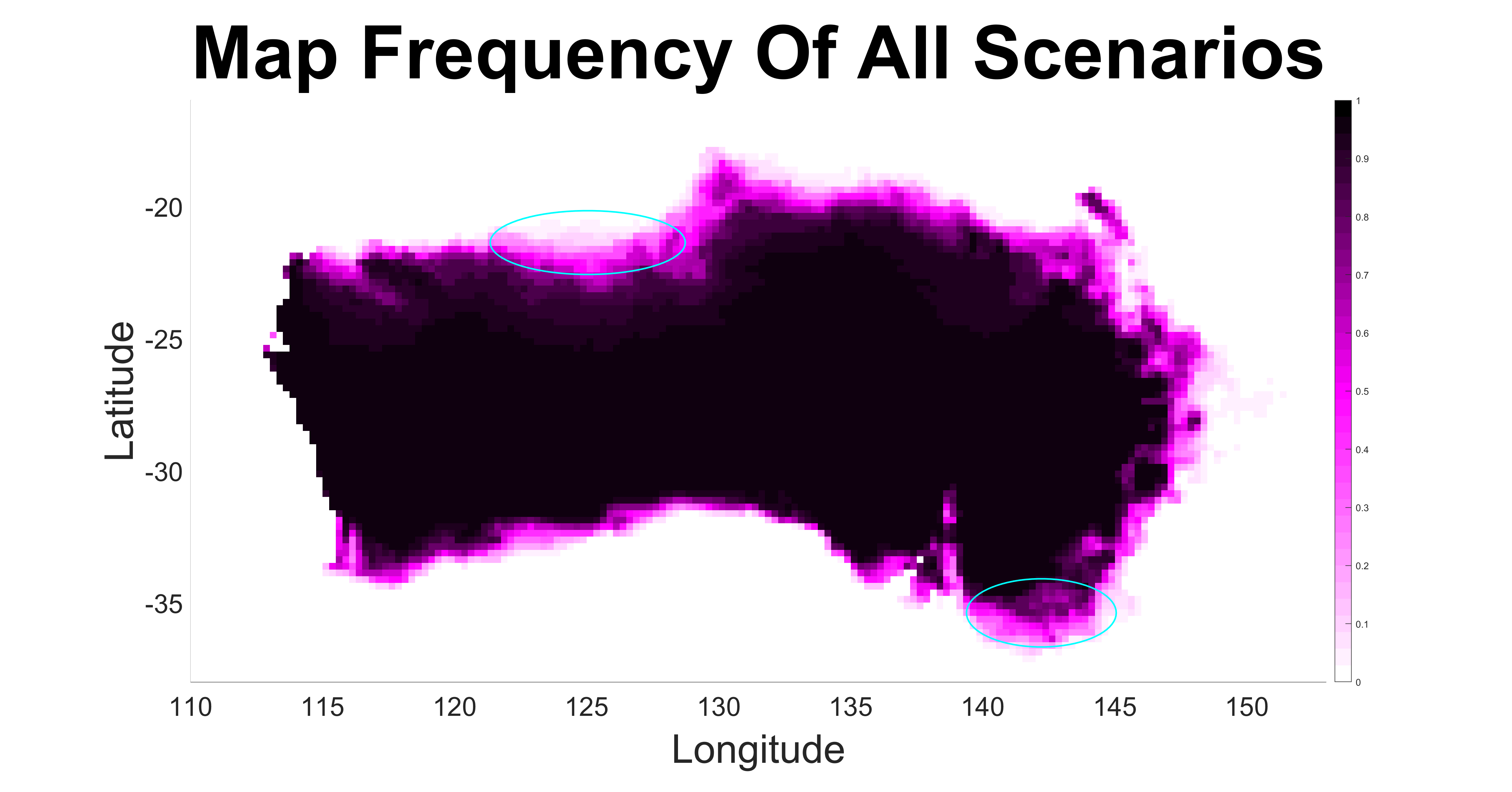}
        \caption{All Scenarios}
        \label{fig:cheetahFrequencies:all} 
    \end{subfigure}
    \begin{subfigure}{.45\textwidth}
        \centering
        \includegraphics[width=\linewidth]{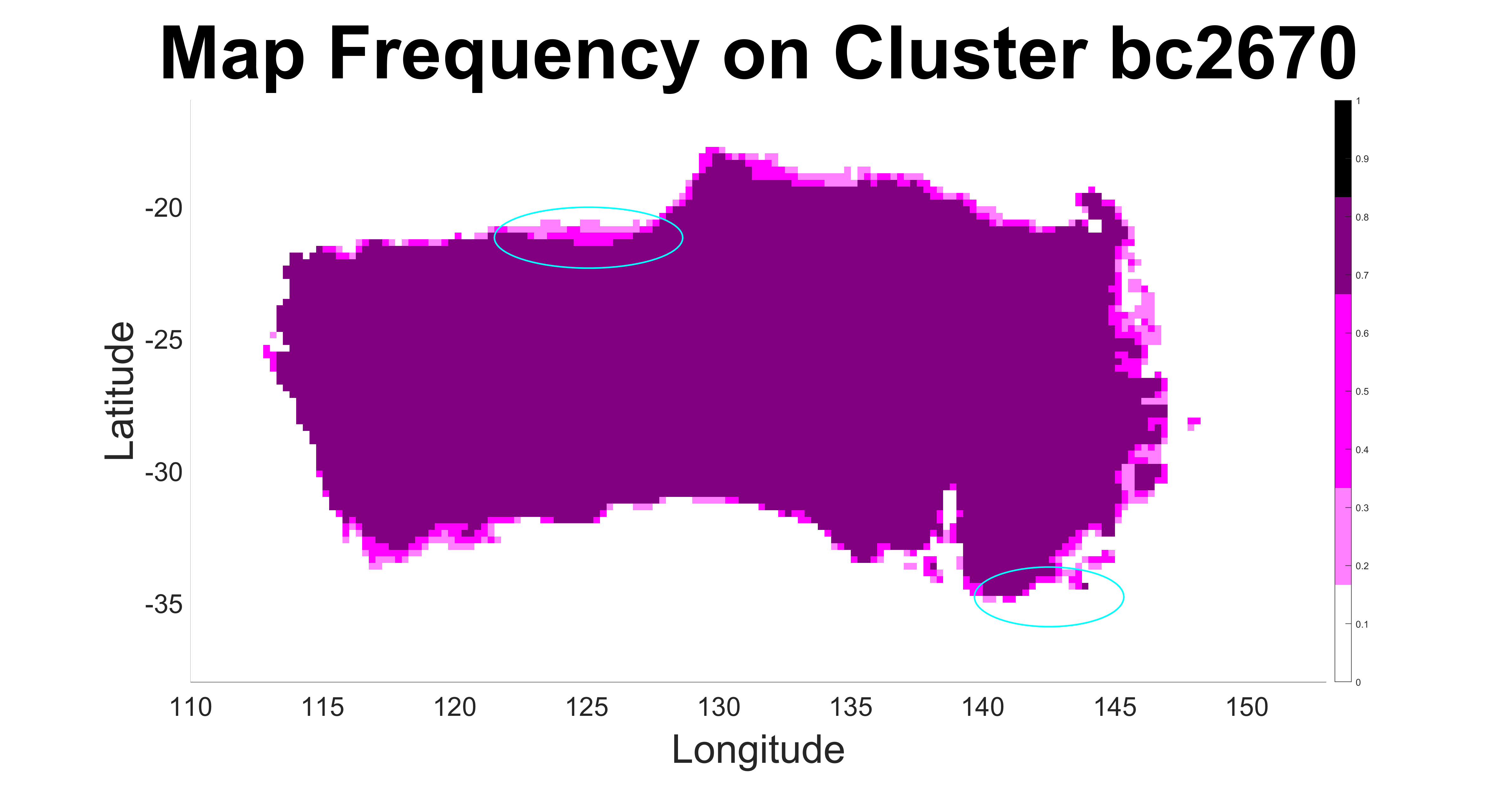}
        \caption{Optimistic Scenarios}
        \label{fig:cheetahFrequencies:optimistic}
    \end{subfigure}
    \begin{subfigure}{.45\textwidth}
        \centering
        \includegraphics[width=\linewidth]{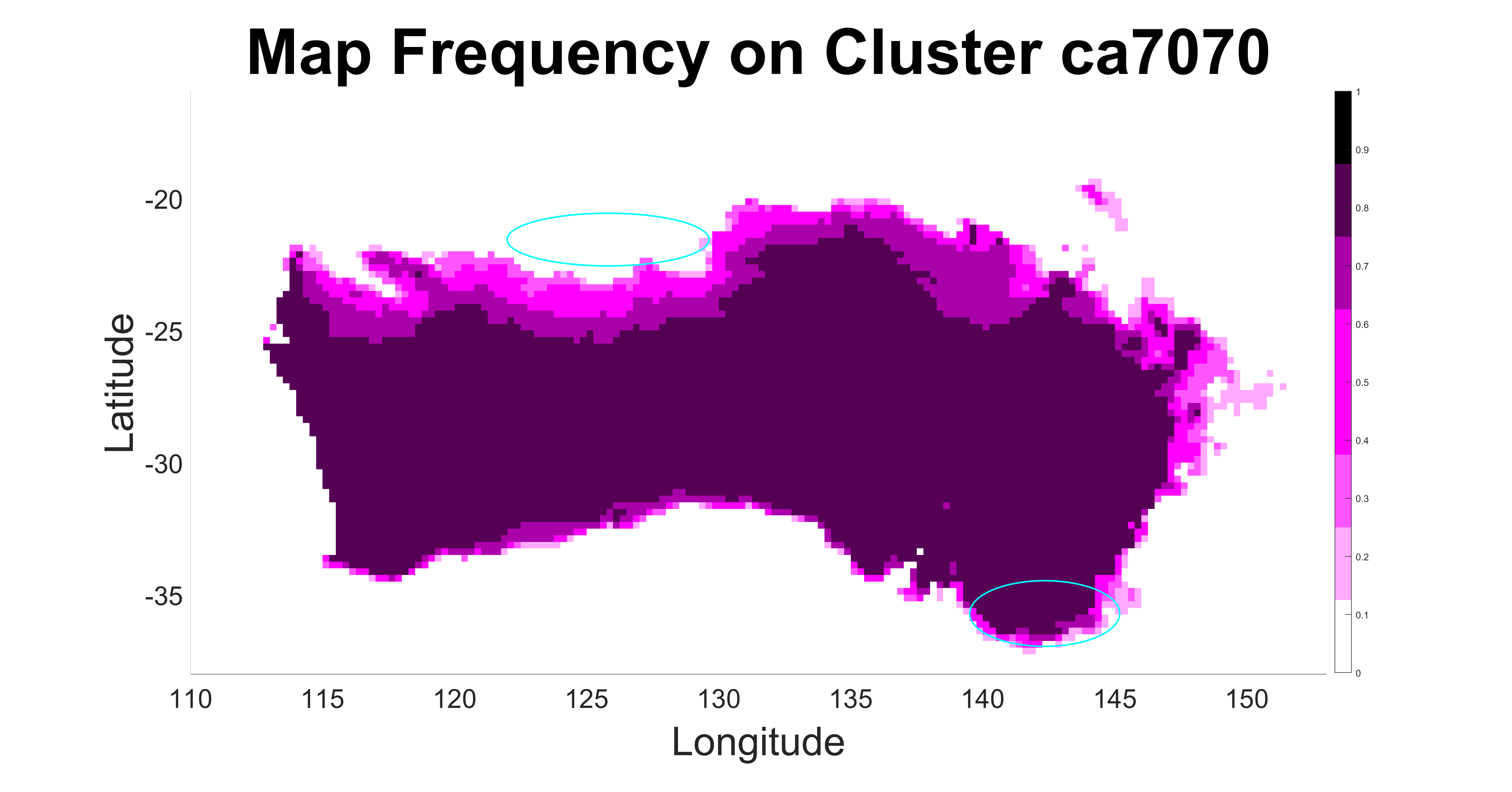}
        \caption{Extreme Scenarios}
        \label{fig:cheetahFrequencies:extreme}
    \end{subfigure}
    \caption{(a) Relative frequency of presences over all scenarios for Australian sandy mouse. A meaningful clustering should have strong similarities within cluster, and differences across cluster. The circled regions show that indeed we have discovered meaningful clusters; there is agreement within clusters in these regions, but differences across clusters. Relative frequency of presences over (b) the ``optimistic cluster'' and (c) ``extreme'' cluster in Figure \ref{fig:mammalClustering:cheetah}}
    \label{fig:cheetahFrequencies} 
\end{figure}

For example, the overlap of all 34 scenario predictions for the Australian mouse is shown in Figure \ref{fig:cheetahFrequencies:all}. We see that the cluster associated with the lowest RCP scenarios (``optimistic" scenarios) accounts for most of the presences in the discrepant regions, whereas the cluster associated with the highest RCP (``extreme" scenarios) accounts for many of the absences (Figure \ref{fig:cheetahFrequencies}). This demonstrates that we have discovered meaningful clusters; there is agreement within clusters but disagreements across clusters. 

\subsection{Individual Climate Variable Clustering} \label{subse4:resulstanddis}
\begin{figure}
    \begin{subfigure}{.5\textwidth}
        \centering
        \includegraphics[width=.9\linewidth]{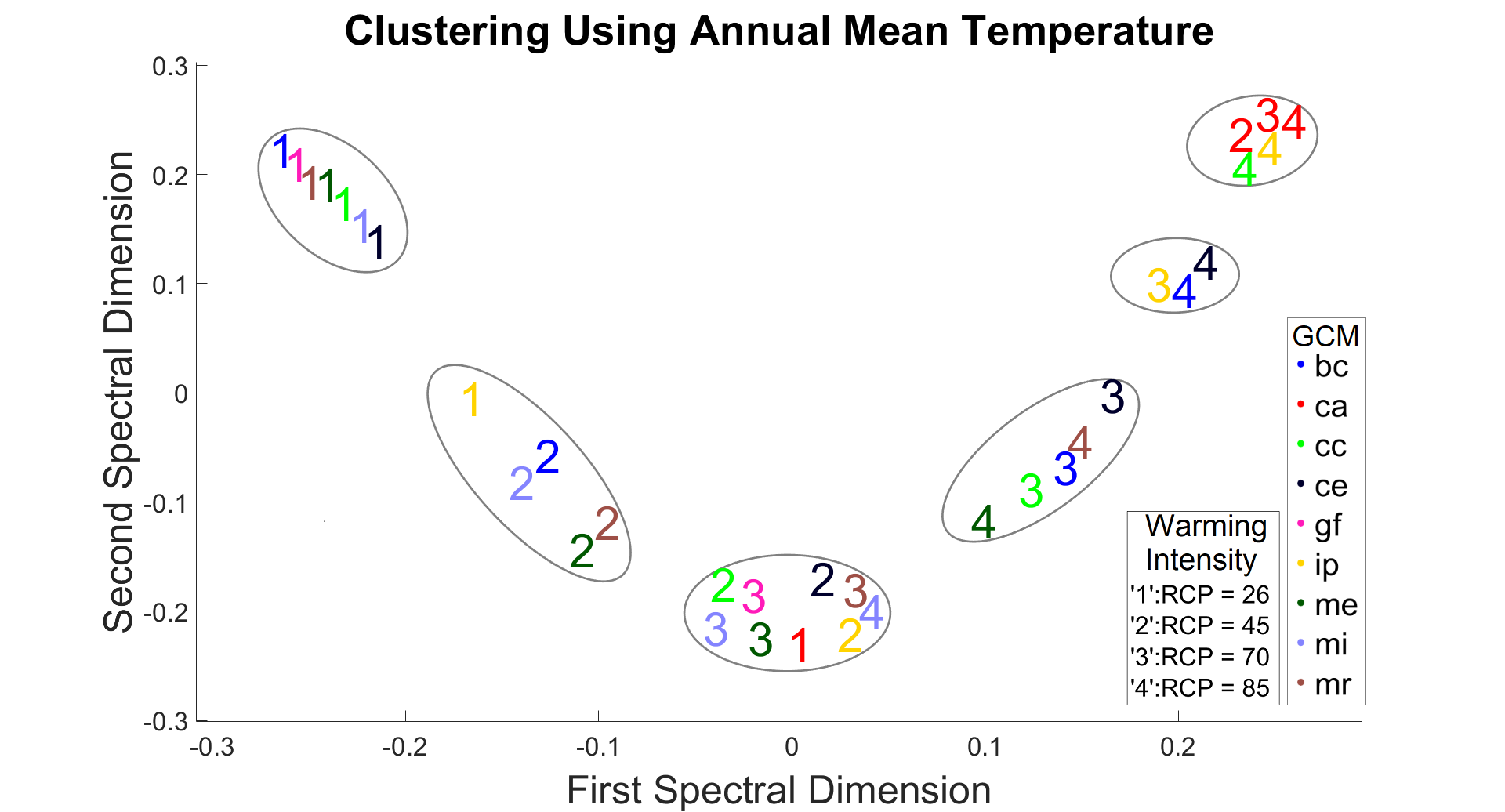}
        \caption{Annual Mean Temperature}
        \label{fig:climateIndividual:temperature}
    \end{subfigure}
    \hfill
    \begin{subfigure}{.5\textwidth}
        \centering
        \includegraphics[width=.9\linewidth]{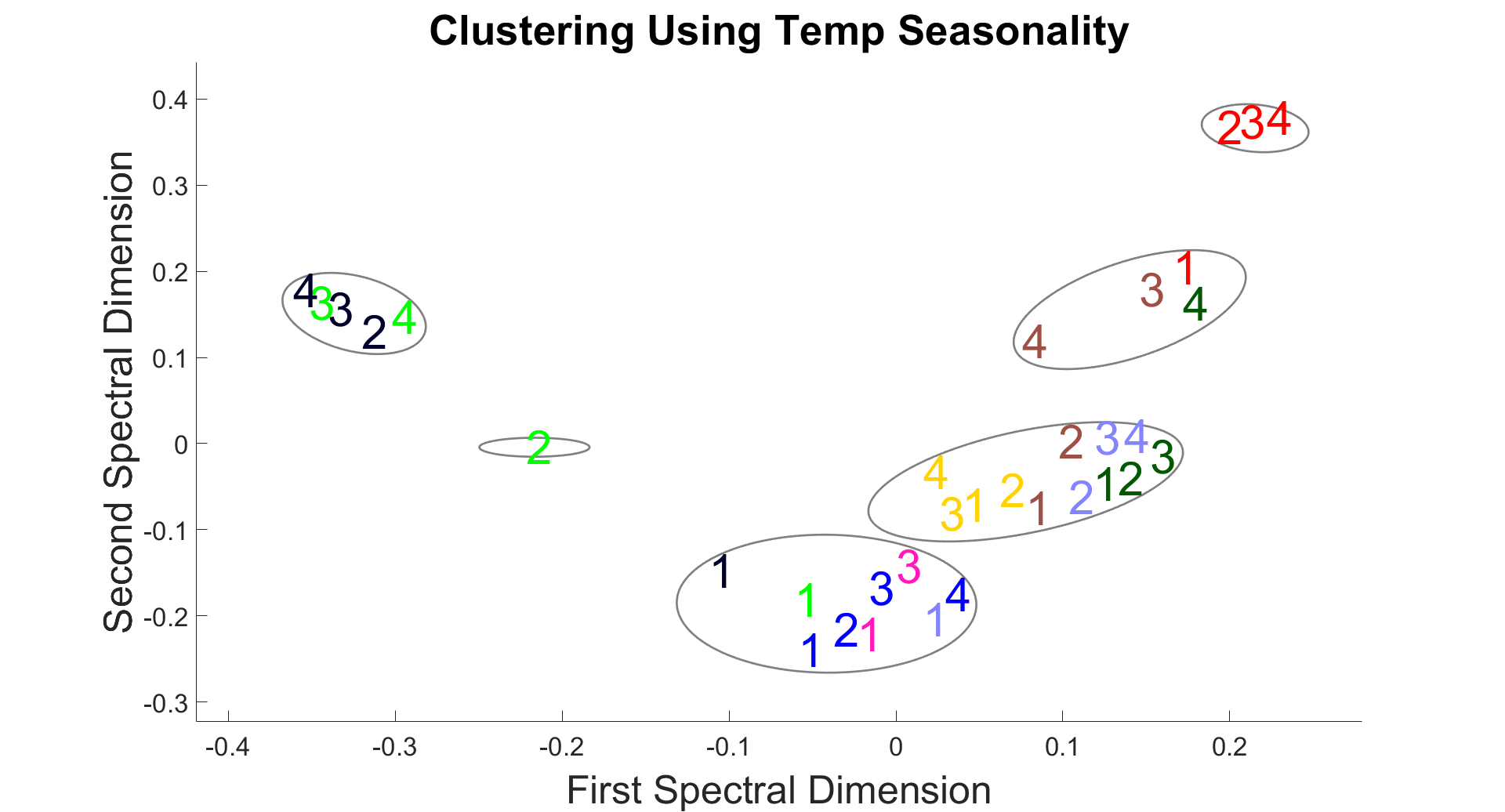}
        \caption{Temperature Seasonality}
        \label{fig:climateIndividual:temperatureSeasonlaity}
    \end{subfigure}
    \hfill
    \begin{subfigure}{.5\textwidth}
        \centering
        \includegraphics[width=.9\linewidth]{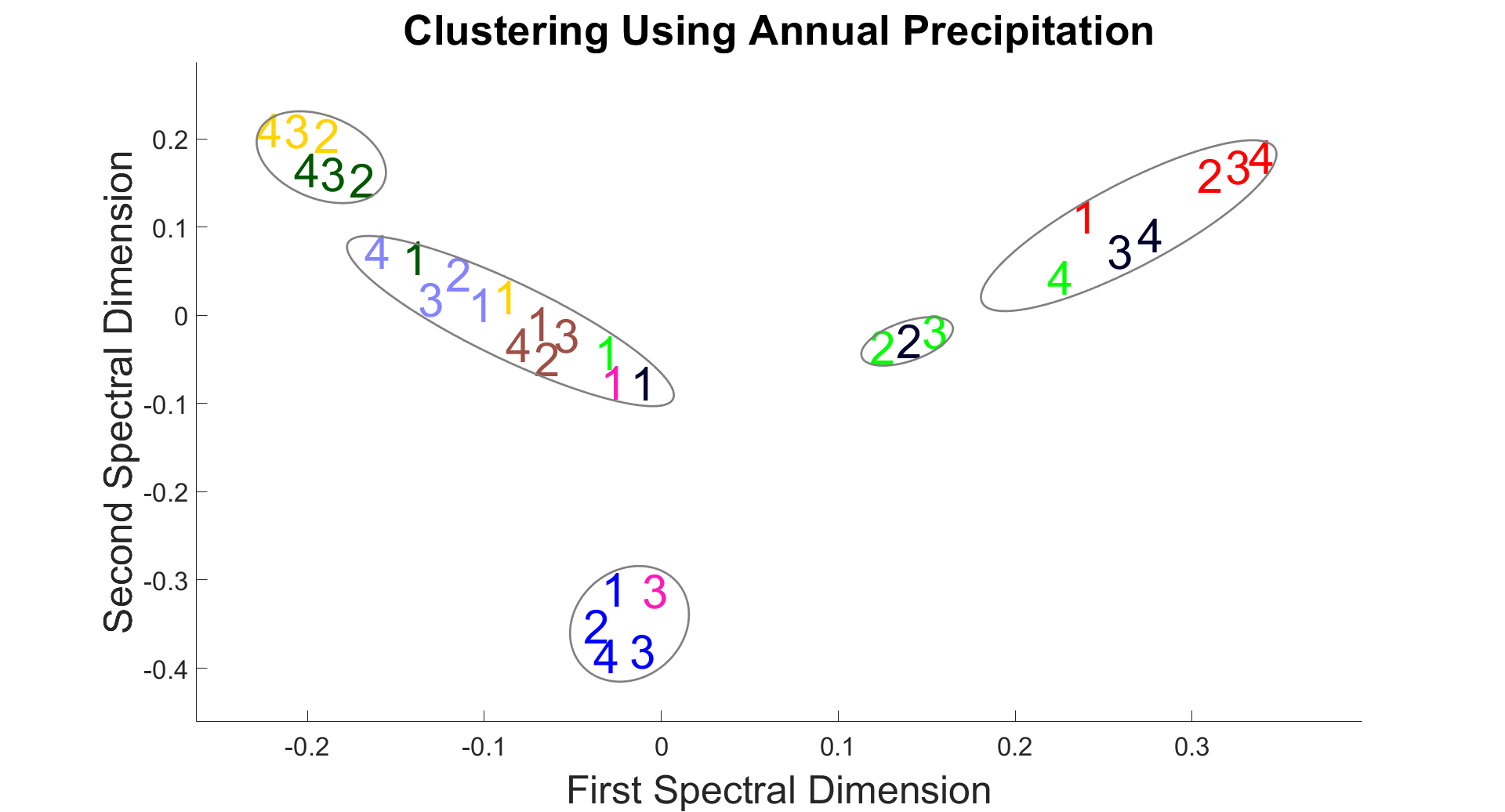}
        \caption{Annual Precipitation}
        \label{fig:climateIndividual:precipitation}
    \end{subfigure}
    \hfill
    \begin{subfigure}{.5\textwidth}
        \centering
        \includegraphics[width=.9\linewidth]{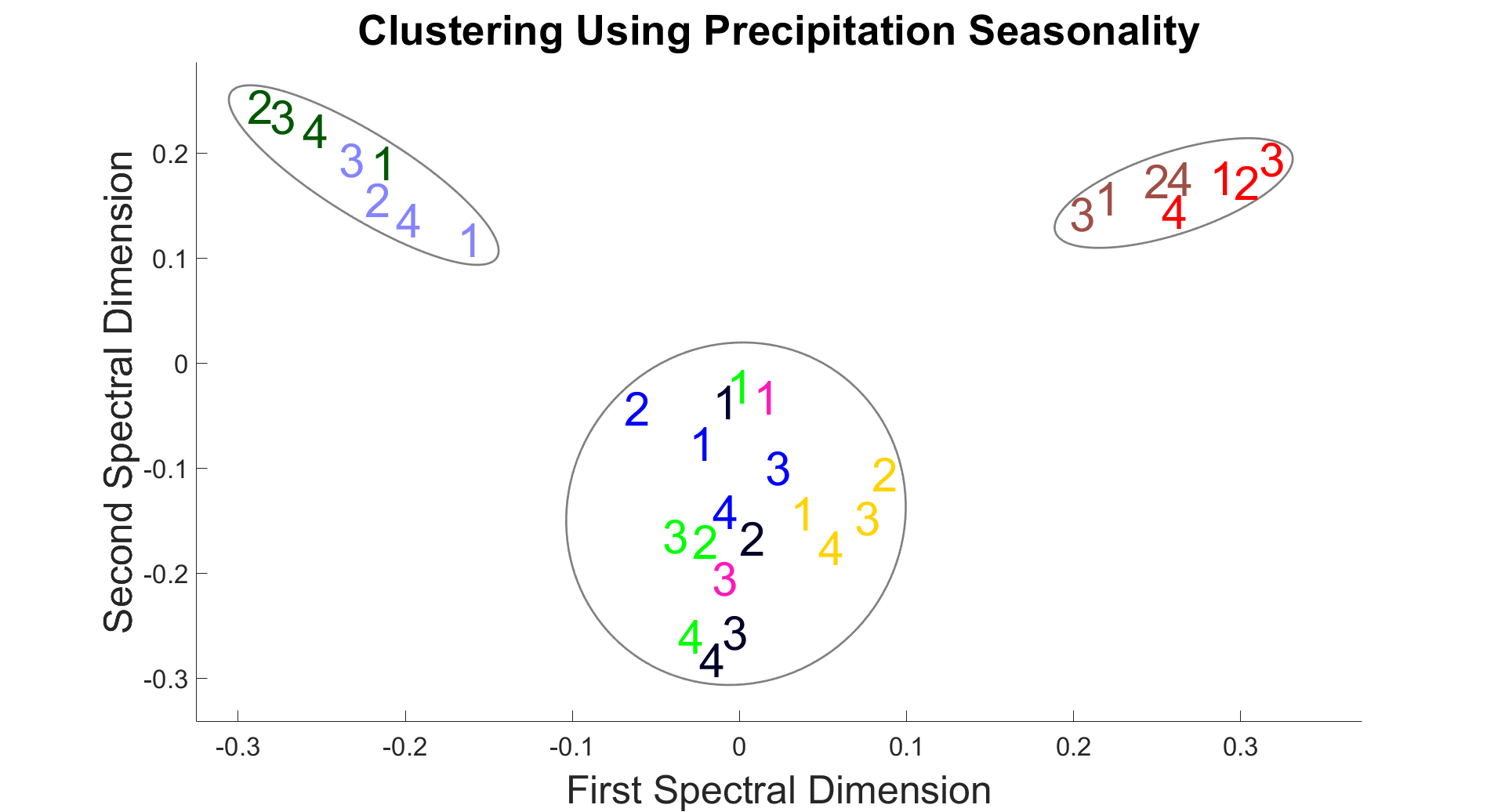}
        \caption{Precipitation Seasonality}
        \label{fig:climateIndividual:precipitationSeasonality}
    \end{subfigure}
    \caption{Spectral clustering of scenarios using individual climate variables (one of the five is not shown due to space). The annual temperature clustering (Figure \ref{fig:climateIndividual:temperature}) is most similar to the ecological clustering of Figure \ref{fig:climateClustering:rangeMap}, clustering mainly by RCP. The other variables cluster mainly by GCM. This is why the climate driven clustering of Figure \ref{fig:climateClustering:climate} is driven mainly by GCM, most of the individual variables cluster by GCM except for temperature. The ecological clustering is important to discern which of these variables are most ecologically relevant, these plots show that annual temperature contains the most ecologically relevant differences between climate models.}
    \label{fig:climateIndividual} 
\end{figure}

We see that the climate-based clustering is meaningfully different than the ecological one, and the clustering is driven mainly by GCM instead of by RCP in the ecological based clustering. The climate-based clustering and ecological one have an adjusted Rand index of only $0.26$, demonstrating their dissimilarity. This is evidence for the importance of considering the specific climate niche occupied by a species in relation to how those conditions are projected to change. Although these same five variables are used to predict the species' ranges, these predicted ranges paint a different picture of the scenario clustering because of how the species are influenced by the climate variables. In fact, we can get a sense of variable importance by clustering based on individual climate variables as opposed to their weighted average, shown in Figure \ref{fig:climateIndividual}. We see that clustering using only the annual temperature creates the most similar clustering to the ecologically driven one (Figure \ref{fig:climateClustering:rangeMap}), suggesting that annual temperature is the most important variable of these five. 

\section{Conclusion} \label{se:conclusion}
We have proposed a novel framework for clustering future scenarios of species range maps. The presented approach is interpretable, flexible, and computationally efficient. We have demonstrated different patterns of clustering depending on the subsets of species: individual species, a subset of species most at risk, and all species. The differences between the climate- and ecological-based clustering highlights the importance of considering species niches; the interaction of climate and ecology is essential to understand the ecologically most important differences between future scenario predictions.

An interesting direction to explore further is to uncover subsets of species that respond differently than others. For instance it may be the case that rodents tend to fare worse under a specific climate model, compared to other mammals. 
A similar area of future research is to determine why some species like the slender treeshew cluster more by GCM instead of the more common pattern of RCP. 
Another extension is to consider how to combine information across species in a more holistic manner as opposed to the simple average. For instance, \cite{dong2013clustering} presents a methodology to cluster according to many graphs, which could be applied to the set of scenario graphs from each species.

\bigskip
\noindent {\bf Acknowledgements}\\
Financial support is gratefully acknowledged from a Xerox PARC Faculty Research Award, National Science Foundation Awards 1455172, 1934985, 1940124, and 1940276, USAID, and Cornell University Atkinson Center for a Sustainable Future. CM acknowledges support from NSF Award HDR-1394790.

% who wrote this?
%Monthly values of minimum temperature, maximum temperature, and precipitation were processed for nine global climate models (GCMs): BCC-CSM2-MR (bc), CNRM-CM6-1, CNRM-ESM2-1, CanESM5, GFDL-ESM4, IPSL-CM6A-LR (ip), MIROC-ES2L, MIROC6 (mi), MRI-ESM2-0 (mr),

\appendix
\section*{Appendix} \label{se:Appendix}
We provide here a table with information about the climate models considered in this work. The name of each model is composed of the respective modeling center/institute and a description of the considered model which is either a Climate System Model (CSM), Climate Model (CM), Earth System Model (ESM) or Earth System (ES) plus a number which indicates the version of the model considered by the respective institute.  
\begin{table}[hbt!]
    \centering
\begin{tabular}{l l l} 
Model & Institute & Abbreviation \\ [0.5ex] 
\hline\hline
BCC-CSM2-MR & Beijing Climate Center, Beijing, China & bc \\ 
CNRM-CM6-1 & Centre National de Recherches Meteorologiques, Toulouse, France & cc  \\
CNRM-ESM2-1 & Centre National de Recherches Meteorologiques, Toulouse, France & ce  \\
CanESM5 & Canadian Centre for Climate Modelling and Analysis, Victoria, Canada & ca \\
GFDL-ESM4 & Geophysical Fluid Dynamics Laboratory, Princeton, USA & ge \\
IPSL-CM6A-LR & Institut Pierre Simon Laplace, Paris, France & ip \\
MIROC-ES2L & Japan Agency for Marine-Earth Science and Technology, & mi \\
& Atmosphere and Ocean Research Institute and National Institute &\\
& for Environmental Studies, Tokyo, Japan & \\
MIROC6 & Japan Agency for Marine-Earth Science and Technology, & me \\
& Atmosphere and Ocean Research Institute and National Institute &\\
& for Environmental Studies, Tokyo, Japan & \\
MRI-ESM2-0 & Meteorological Research Institute, Tsukuba, Japan & mr
\end{tabular}
\caption{\label{tab:ScenTable} Institutes and abbreviations of models participating CMIP. More details can be found in \href{https://www.wcrp-climate.org/modelling-wgcm-mip-catalogue/modelling-wgcm-cmip6-endorsed-mips}{CMIP6}.}
\end{table}

\clearpage

\bibliographystyle{apalike}
\bibliography{bibliography}
\end{document}